\documentclass[11pt]{article}
\usepackage[margin=1in]{geometry}
\usepackage{amsthm,amsmath,amsfonts,amssymb}
\usepackage{graphicx}
\usepackage{booktabs}
\usepackage{float}
\usepackage{stfloats}
\usepackage{xurl}
\usepackage[authoryear,round]{natbib}
\usepackage{tikz}
\usepackage{tikz-3dplot}
\usetikzlibrary{3d,arrows.meta,positioning,calc,patterns,decorations.pathreplacing}
\usepackage{pgfplots}\pgfplotsset{compat=1.18}
\usepackage{authblk}
\usepackage[colorlinks=true,citecolor=blue,urlcolor=blue]{hyperref}
\usepackage[section]{placeins}
\usepackage{caption}
\theoremstyle{definition}
\newtheorem{definition}{Definition}

\begin{document}

\title{The Quantification of Replicability}
\author[1]{Ester Alongi\thanks{ester.alongi@phd.unipd.it}}
\author[2]{Gianmarco Alto\`e\thanks{gianmarco.altoe@unipd.it}}
\author[3]{Giovanni Parmigiani\thanks{gp@ds.dfci.harvard.edu}}
\affil[1]{Department of Statistical Sciences, University of Padua, Padua}
\affil[2]{Department of Developmental Psychology and Socialisation, University of Padua, Padua}
\affil[3]{Department of Data Science, Dana-Farber Cancer Institute, Boston \& Department of Biostatistics, Harvard T.H. Chan School of Public Health, Boston}
\date{\today}
\maketitle
\begin{abstract}
What does it mean to measure the replicability of scientific results? Is it merely determining whether results are consistent across studies? Much of the existing literature on replicability draws conclusions exclusively from the comparison of study results, in terms of observed estimates, uses a variety of different metrics, and ultimately collapses the comparison into a binary verdict of success or failure. This fails to differentiate between repeated errors and corroboration of the shared generative effect. We develop a unified Bayesian framework in which distinct but related notions of replicability are formally defined and jointly embedded within a single inferential structure. A single model answers four complementary questions: agreement among independent studies; consistency between the individual studies and the shared generative effect; agreement between an existing body of evidence and a new study from the same super-population; inference on the shared generative effect. In practice, we suggest quantifying replicability through posterior probabilities relative to regions of practical relevance. We illustrate these ideas in an eQTL (expression quantitative trait locus) analysis assessing the replicability of a regulatory effect across human ancestral groups.
\end{abstract}


\section{Introduction}\label{sec:introduction}

\subsection{Preface}\label{sec:preface}
Replicability is central to science: the robustness of a result is an essential element before a scientific claim can be made. Yet, despite its centrality, it is assessed in many different and empirical ways that capture only partial information about the phenomenon of interest. In this paper, we offer a unified conceptual framework for discussing why, when and how to quantify replicability.

\subsection{The Crisis of Replicability}
In an influential paper, \citet{Ioannidis2005} argued that the fraction of false positive findings in many areas of science is too high. In the past decade, systematic, large-scale projects have set out to assess this empirically, and their results have raised widespread concern. \citet{open2015estimating} attempted to replicate one hundred studies published in 2008 in three high-ranking psychology journals and found that, while $97\%$ of the original studies reported statistically significant effects, only $36\%$ of the replications did, with replication effect sizes roughly half the originals on average. A parallel effort in preclinical cancer biology reached a comparable conclusion: \citet{Errington2021} found that replication effect sizes were substantially smaller than the originals, with a median replication effect size $85\%$ smaller, and that positive results were considerably less likely to replicate than null ones ($40\%$ of $97$ positive vs.\ $80\%$ of $15$ null effects evaluated across five criteria).
A larger project found the same pattern: \citet{tyner2026investigating} attempted to replicate $274$ claims from $164$ papers across $54$ social and behavioral science journals, finding that $151$ of $274$ replications ($55.1\%$) yielded a significant result in the original direction.
 
Beyond the empirical findings of the large-scale projects just illustrated, however, these projects exposed an obstacle on a deeper level: there is no single answer to what it means for a result to have replicated, and the rate one reports depends heavily on the criterion adopted. Thus, the crisis is not only empirical but conceptual, and it is the conceptual question that concerns us here.

\subsection{State of the Art}\label{sec:sota}
In 2019, the U.S. National Academies of Sciences, Engineering, and Medicine published a report on Reproducibility and Replicability in Science, defining replicability as ``obtaining consistent results across studies aimed at answering the same scientific question, each of which has obtained its own data'' \citep{NAS2019}. This conceptual definition provides a baseline, yet translating this definition into statistical practice has not produced consensus but a proliferation of methods: a recent scoping review by \citet{heyard2025scoping} makes the scale of this fragmentation explicit, cataloging $50$ distinct approaches to quantifying replicability, $37$ of them based on explicit statistical models or formulas, and documenting a corresponding lack of methodological consensus. Computing these metrics requires many different models: significance tests on the direction or magnitude of the effect \citep{Errington2021, open2015estimating}, checks of whether the replication effect falls within a prediction interval of the original \citep{patil2016should}, or Bayes factors comparing the two studies \citep{verhagen2014bayesian}. These approaches rest on different assumptions about the data-generating mechanism and between-study heterogeneity, so reliance on multiple, mutually inconsistent models can lead to divergent conclusions about the same evidence. \citet{camerer2016evaluating} and \citet{tyner2026investigating} show this concretely. \citet{camerer2016evaluating} replicated 18 studies published in the American Economic Review and the Quarterly Journal of Economics between 2011 and 2014, judging replication against four indicators and obtaining replicability rates that range from 61.1\% to 83.3\%. \citet{tyner2026investigating} applied thirteen alternative measures of replication success to a single set of studies and obtained success rates ranging from 28.6\% to 74.8\%. Therefore, whether a body of findings is judged to replicate depends substantially on the metric chosen.

A further feature of this landscape concerns not the metrics themselves but how they are used, in two distinct respects: which hypothesis is being tested, and how the answer is reported. Most metrics rest on a point hypothesis, either the null itself or a shifted point value, rather than on a region of practically equivalent effects. And while most metrics can in principle be reported continuously (in the scoping review of \citet{heyard2025scoping}, 47 of the 50 can quantify reproducibility on a continuous scale), in practice the continuous evidence is very often collapsed into a binary verdict of success or failure.
An example of both point hypothesis and collapse of the response to a binary result is the agreement on statistical significance, that is, whether both studies reached significance and in the same direction \citep{Errington2021}. Two replication attempts with nearly identical effect estimates can receive opposite verdicts when one falls just below and the other just above the significance cutoff, so that an arbitrary boundary on a continuous statistic, and not any real difference in the evidence, decides whether a finding is judged to replicate. The sceptical $p$-value \citep{held2020new} moves the null away from zero, yet it is again used as a dichotomous criterion for replication success \citep{micheloud2023assessing}. The same dichotomizing logic of the response underlies several other criteria in common use: agreement in the sign of the effect \citep{Errington2021}; inclusion of the original estimate within the $95\%$ confidence interval of the replication effect \citep{cova2021estimating}; or, where meta-analytic tools are used at all, merely whether the meta-analytic estimate is significant \citep{open2015estimating}. Bayesian metrics, which improve on this in principle by using probability distributions, retain point hypotheses: replication Bayes factors weigh a point null hypothesis against an alternative given by the original study's posterior \citep{verhagen2014bayesian}, so that although the evidence is graded, the comparison is between the presence and the absence of an effect rather than between degrees of practical relevance. 

Some large-scale projects characterize the observed variability with Bayesian meta-analytic models, without defining a criterion for the success of replication \citep{coretta2023multidimensional}, describe replicability by using descriptive summaries \citep{huntington2021influence} or assess replicability through prediction markets \citep{botvinik2020variability} and expert surveys \citep{alipourfard2021systematizing}.
 
Most of these methods draw their conclusions from the comparison of observed estimates, and thus remain confined to the empirical space. This has two consequences. First, because the verdict rests on estimates, a single underpowered study can drive it: a study with insufficient power to detect a real effect may be judged not to replicate for reasons that have nothing to do with the phenomenon under study, and a metric that stays at the data level cannot separate this from true non-replication. Second, agreement between two estimates does not tell us whether the studies captured the same real effect or just repeated the same error, because the comparison never looks at the underlying shared generative effect. Even when the generative structure is present, for example through a hierarchical structure \citep{pawel2020probabilistic}, it tends to serve only to model between-study heterogeneity rather than to quantify replicability itself: what is missing is the use of that generative structure to ask whether the study-specific effects and the shared generative effect jointly support a coherent conclusion about the effect of interest. 

A parameter-level definition of replicability does exist in the multiple-testing literature, where the claim of interest is that the study-specific parameters are non-null in a common direction in at least $k$ of the $S$ studies \citep{benjamini2008screening}, with an empirical-Bayes counterpart in genomics \citep{heller2014replicability}; see \citet{bogomolov2023replicability} for a review.

These three aspects, whether a metric poses replicability as a point hypothesis, whether it collapses the answer into a binary verdict, and whether it remains confined to the empirical space, are mutually independent. The replication Bayes factor tests a point hypothesis yet reports continuous evidence rather than a binary verdict \citep{verhagen2014bayesian}, so a point hypothesis does not force a binary answer. A posterior predictive check reports a continuous quantity yet, by asking whether a new study estimate is compatible with the original one, can stay in the space of observable estimates \citep{pawel2020probabilistic}, so a continuous answer does not by itself reach the parameter level. Conversely, the partial-conjunction approach of \citet{bogomolov2023replicability} operates at the parameter level yet still tests a point hypothesis on each effect, its sign, or its exceeding a threshold, rather than a region of practical relevance, so working at the parameter level does not by itself replace the point hypothesis with a region.

\subsection{The Added Value of Replicability}
\begin{figure}[ht]
\centering
\includegraphics[width=0.45\textwidth]{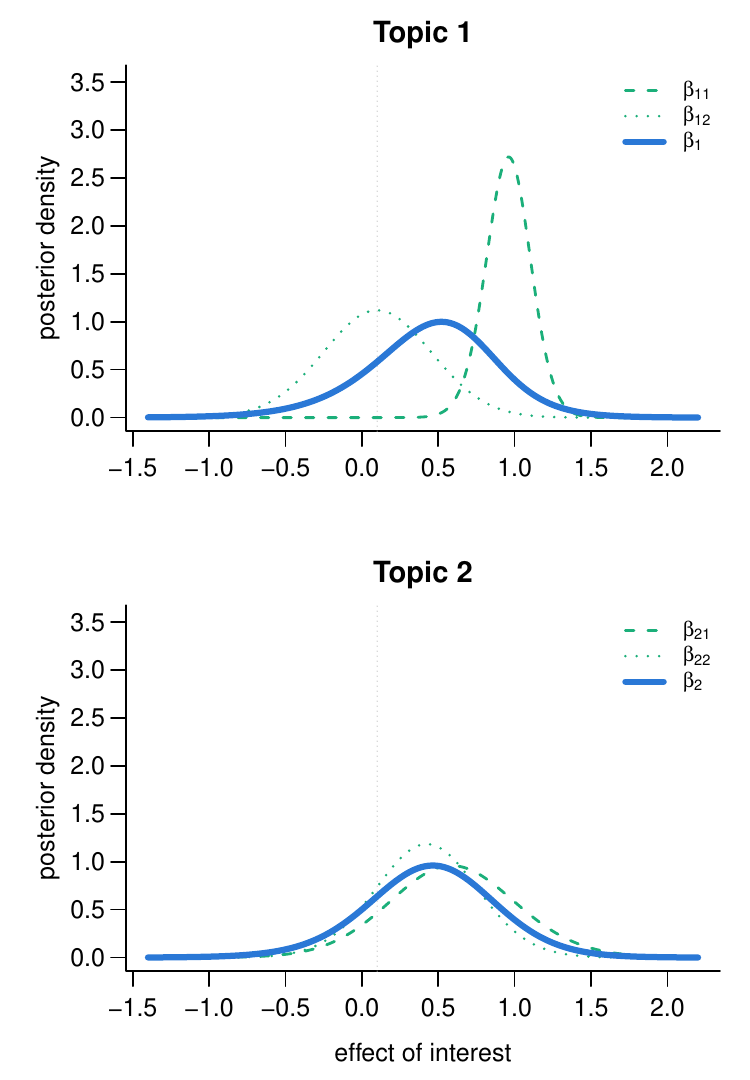}
\caption{Posterior densities of the study-specific effects
$\beta_{11}$, $\beta_{12}$, $\beta_{21}$, $\beta_{22}$ (green, thin, distinguished by dashing) and the shared generative effects $\beta_1$ and $\beta_2$ (blue, thick solid); the dotted vertical line marks $\varepsilon=0.1$.}
\label{fig:motiv}
\end{figure}

A simple example can make concrete the challenges this paper addresses. Figure~\ref{fig:motiv} illustrates two topics, each a pair of studies of equal size (25 observations in every study) addressing the same research question: whether some effect of interest is above a specified threshold $\varepsilon$, shown by the vertical dotted line.

The between-study standard deviations $\tau_1$ for topic 1 and $\tau_2$ for topic 2 are estimated separately within each topic, but with the same diffuse prior $\tau_1, \tau_2 \stackrel{\text{iid}}{\sim} \mathrm{LogNormal}(\log 0.5,\,0.50^2)$. The residual variances $\sigma^2_{11}$, $\sigma^2_{12}$, $\sigma^2_{21}$ and $\sigma^2_{22}$ are also unknown, with the sample variances $s^2_{11}$, $s^2_{12}$, $s^2_{21}$, $s^2_{22}$ entering the sufficient statistics. Effects are assessed relative to a specified threshold $\varepsilon = 0.1$: in the last column of Table~\ref{tab:s3} we report the posterior probability that the effect is above the threshold $\varepsilon$, $\Pr(\beta > \varepsilon \mid y)$. The two topic 1 studies differ sharply in precision: the first study is tight ($s^2_{11}=0.50$) and the second one is noisy ($s^2_{12}=4.0$), an eight-fold ratio. By contrast, the topic 2 studies are both noisy ($s^2_{21}=8.67$ and $s^2_{22}=4.33$). In Table~\ref{tab:s3} we report all the sufficient statistics and the prior and posterior summaries of each parameter.

\begin{table}[!ht]
\caption{Sufficient statistics ($\bar y$ and $s^2$), prior and posterior summaries, posterior probabilities that the effects are above the threshold $\varepsilon$.}
\label{tab:s3}
\centering\footnotesize
\setlength{\tabcolsep}{6pt}
\begin{tabular}{lccccccc}
\toprule
 & \multicolumn{2}{c}{data} & \multicolumn{2}{c}{prior} & \multicolumn{2}{c}{posterior} & \\
\cmidrule(lr){2-3}\cmidrule(lr){4-5}\cmidrule(lr){6-7}
param.\ & $\bar y$ & $s^2$ & mean & sd & mean & sd & $\Pr(>\varepsilon\mid y)$\\
\midrule
$\beta_{11}$    & $1.0$  & $0.50$ & $0$ & $1.19$ & $0.96$ & $0.15$ & $1.0$\\
$\beta_{12}$    & $-0.11$ & $4.0$ & $0$ & $1.19$ & $0.099$ & $0.36$ & $0.50$\\
$\beta_{1}$     & --- & --- & $0$ & $1.0$    & $0.45$ & $0.44$ & $0.81$\\
$\tau_{1}$      & --- & --- & $0.57$ & $0.30$ & $0.60$ & $0.26$ & ---\\
$\sigma^2_{11}$ & --- & $0.50$ & $1.0$ & $1.0$    & $0.57$ & $0.158$ & ---\\
$\sigma^2_{12}$ & --- & $4.0$ & $1.0$ & $1.0$    & $3.59$ & $0.99$ & ---\\
\midrule
$\beta_{21}$    & $0.79$  & $8.67$ & $0$ & $1.19$ & $0.60$ & $0.43$ & $0.88$\\
$\beta_{22}$    & $0.39$  & $4.33$ & $0$ & $1.19$ & $0.42$ & $0.34$ & $0.83$\\
$\beta_{2}$     & --- & --- & $0$ & $1.0$    & $0.45$ & $0.44$ & $0.80$\\
$\tau_{2}$      & --- & --- & $0.57$ & $0.30$ & $0.49$ & $0.24$ & ---\\
$\sigma^2_{21}$ & --- & $8.67$ & $1.0$ & $1.0$    & $7.49$ & $2.041$ & ---\\
$\sigma^2_{22}$ & --- & $4.33$ & $1.0$ & $1.0$    & $3.82$ & $1.04$ & ---\\
\bottomrule
\end{tabular}
\end{table}

Looking only at the study-specific distributions, the two situations appear very different: in topic 1 we have a tall, tight posterior decisively above $\varepsilon$ ($\Pr(\beta_{11}>\varepsilon\mid y)=1.0$) next to a low, wide posterior ($\Pr(\beta_{12}>\varepsilon\mid y)=0.50$) straddling it, while in topic 2 the posterior distributions of the two effects are nearly identical ($\Pr(\beta_{21}>\varepsilon\mid y)=0.88$ and $\Pr(\beta_{22}>\varepsilon\mid y)=0.83$). Thus, from the point of view of a replicability analysis, one could conclude that topic 1 does not replicate while topic 2 does. However, the posterior distributions of the shared generative effects $\beta_1$ and $\beta_2$ are very similar between the two topics; indeed, the two probabilities that the effect is above the threshold $\varepsilon$ are nearly equal, $\Pr(\beta_{1}>\varepsilon\mid y)=0.81$ and $\Pr(\beta_{2}>\varepsilon\mid y)=0.80$. From a meta-analysis point of view, one could thus conclude that the two topics contain the same scientific information about the effect.

Therefore, if we ran only one of these two analyses, we would lose some important information. If we saw only the replicability we would lose the information that from an inferential point of view we would have strong evidence for the effect in both topics. At the same time, if we saw only the meta-analysis, we would lose information about how the conclusions should apply in the real world: in topic 1, the conclusion is reached from studies that point to different magnitudes of the effect, while in the second, the studies point to the same magnitude of the effect. The two carry different implications for how the effect might behave in practice. From the first case, one might infer that further investigation is needed to assess whether the effect is stable between different studies. From the second, one might reasonably expect future studies to agree, and read the finding as robust well beyond the samples at hand.

Like the meta-analysis, an analysis of heterogeneity by itself would also lead to a misleading conclusion: knowing only the information about $\tau$, we would conclude that there is no important difference in terms of heterogeneity between the two topics (the posterior means of the between-study standard deviations $\tau_1$ and $\tau_2$ for the two topics are $0.60$ and $0.49$, respectively), so that the hypothesis that the results generalize to the entire super-population would not seem entirely unreasonable. By contrast, a replicability analysis brings to light that, even though the heterogeneity and the shared generative effects are similar across the two topics, the situations can be markedly different in terms of concordance; this could make the researcher aware of the difference and prompt them to explore the reasons why there is this lack of replicability. Understanding whether the lack of replicability is due to a true high heterogeneity in the super-population or a limitation in one of the studies helps decide whether the results may not be generalizable to the whole super-population or whether the analysis needs further investigation through other studies. For example, the posterior mean of the second study in topic 1 barely departs from the prior mean of zero ($\beta_{12}$ has posterior mean $0.099$) and its posterior is very wide (posterior standard deviation $0.36$, against $0.15$ for $\beta_{11}$); this points to an imprecise, possibly unreliable measurement in that study rather than to genuine between-study heterogeneity. Indeed, understanding the sources of non-replicability can help determine whether and how to plan a new study. If it stems from a problem in one study, a further study is needed to establish whether the effect truly replicates, since the current verdict is unreliable; if instead it reflects true heterogeneity in the super-population of interest, the non-replication is itself a real and informative finding: rather than re-running the study to confirm it, one can account for the heterogeneity explicitly in the model and conclude that the results do not generalize across the whole super-population.

A replicability analysis thus adds another layer to our understanding of a phenomenon. One might think that, from a Bayesian point of view, the posterior of the shared generative effect summarizes all the uncertainty about the phenomenon. The posterior of the shared generative effect quantifies how much uncertainty remains about the average effect, and the posterior heterogeneity $\tau$ captures how much uncertainty remains about the differences between studies. A replicability assessment asks something further: whether the evidence is coherent enough that this average reflects a stable underlying phenomenon rather than the coincidence of discordant results.
Therefore, in our view, a replicability analysis should take into account all these nuances, separating evidence that the effect is above a specified threshold from evidence for its replicability, and then weighing the link between them.

\subsection{Contribution}\label{sec:contribution}
The example just presented reveals the added value of a replicability analysis; in this section, we outline how we develop it throughout the paper. 

We organize our proposal around three questions that arise from the replicability definition given by \citet{NAS2019}. First, what must the studies share to aim at the same scientific question? Second, how do we take the same scientific question into account within a statistical model? And third, given such a model, how do we formalize what it means for results to be ``consistent''?
 
Consider the first question. If two independent studies investigate the same scientific problem despite differing in context or sample, one way to structure their comparison is to posit a shared, underlying super-population from which both study-specific effects derive. When the phenomenon is fundamentally the same, it is reasonable to assume that the study-specific effects have a shared generative effect, even as they deviate locally through study-specific characteristics. This shared generative effect aligns with how \citet{Nosek2020} characterize replication in practice: not the repetition of procedural steps, but the conduct of a new study for which ``any outcome would be considered diagnostic evidence about a claim from prior research.'' A study can be diagnostic about a previous one when the two share an underlying effect to be diagnostic about, which is what the super-population supplies.
 
This answers the second question in outline: to carry the shared scientific question into the model, we adopt a hierarchical formulation in which a shared generative effect represents the super-population from which the studies are drawn.

As for the third question, how to measure replicability, we propose to assess the replicability of the underlying continuous effects of interest, not that of the statistics that summarize them. We read a family of metrics directly from the model's parameters. Different notions of replicability are jointly embedded within a single inferential structure, and replicability is evaluated as a posterior probability relative to a prespecified threshold $\varepsilon$. A dichotomization of the replicability \emph{question} remains, since we ask whether an effect is relevant with respect to a threshold, but the threshold delimits a region of the parameter space rather than a point, and the \emph{answer} is not dichotomized: it is reported as a posterior probability of that region rather than as a verdict. The threshold enters only when forming the event whose posterior probability is reported, not in the effects it estimates, so the continuous effects themselves are never collapsed. Finally, the hierarchical generative structure carries not only the shared generative effect and its heterogeneity, but also the information needed to quantify replicability itself.

\section{Replicability: Concepts and Terminology}\label{sec:glossary}
In this section we introduce some terminology that helps clarify when one study can be considered a replication attempt of another.

\subsection{Fixed and Random Factors}\label{sec:fixedandrandomfactors}
To describe the components of a study we use the UTOS scheme of
\citet{alma990026209230107871}:
\begin{itemize}
  \item \emph{Units}: the elements of the target population from which the sample is drawn.
  \item \emph{Treatments}: the independent variables.
  \item \emph{Observations}: the operations through which the outcome is measured.
  \item \emph{Setting}: the environmental or contextual factors.
\end{itemize}
As an illustration, in a study that estimates the effect of a treatment on some outcome, the units are the individuals enrolled in the study, the treatment is the intervention whose effect is under investigation, the observations are the operations
by which the response is measured in each unit, and the setting comprises the conditions and procedures under which the study is carried out.

Each of these components may be classified as a fixed or a random factor \citep{machery2020replication}. A random factor is a sample drawn probabilistically from a broader target population; a fixed factor is not sampled but set deliberately by the research design.

Typically, the treatment, the observations, and the setting are fixed, while the units are random, being sampled from a target population. Whether the target population is itself fixed or random distinguishes two situations. If the target population is fixed, the units are the only random factor, and the parameter of interest is a single quantity shared across studies. If instead the target population is random, so that the studies arise from a common super-population but correspond to different target populations within it, the parameter of interest is study-specific. This distinction, whether two studies sample from the same target population or from different ones, organizes the rest of the section.

\subsection{Replication and Extension}\label{sec:replication_extension}
A \emph{replication} \citep{machery2020replication} repeats a study while resampling its random factors from the same target population, holding the fixed factors constant. The parameter of interest can therefore be regarded as theoretically identical from one study to the next. 
The aim of such a study is to evaluate what \citet{machery2020replication} calls \emph{reliability}: an experiment is reliable if, on repeatedly resampling new values for the random factors while holding everything else constant, a stable finding would recur with high frequency. Resampling therefore probes the reliability of the conclusion about the effect of interest: by assessing agreement across new samples, we confirm that the relationship first observed is a truly replicable phenomenon.

When the second study instead samples from a different target population, or when it alters one or more of the fixed factors, it is not a replication, but an \emph{extension} \citep{machery2020replication}.
Rather than testing reliability, an extension assesses \emph{validity} as well as the \emph{invariance range} of a phenomenon: if the studies sample from different target populations, each study then has its own effect and so in the extension we ask whether the conclusions about the effect of interest are robust across contexts.

\subsection{A Continuum of Replication}\label{continuumofreplication}
For \citet{machery2020replication}, whether the second study is a replication or an extension hangs on the single question of whether it samples from the same target population as the original. We retain his criterion but read it as a matter of degree rather than a dichotomy. When the between-study heterogeneity from which the studies are generated is null, the studies derive from the same target population: we are in the case of a replication. As the generative between-study heterogeneity increases, the studies can no longer be regarded as exchangeable draws from the same target population, and we are in the case of an extension. Intermediate levels of the heterogeneity describe target populations that are related but not identical, the graded regime between Machery's two categories, so that replication and extension appear as the two ends of a single continuum. How this spectrum is governed by the heterogeneity parameters is made precise in Section~\ref{sec:model}, once the model is introduced.

In practice, it is not possible to establish with certainty whether we are dealing with a replication or an extension, since we cannot know the true value of the between-study heterogeneity. Consider, for example, a family of drugs that all share the same putative mechanism of action, all thought to inhibit a given molecule. The first study tests the marketed drug and the others test later variants. We want to understand whether the effect found in the first study is stable across the others. But whether we are asking about the reliability of the original effect or about its invariance across studies of different target populations is not obvious: the variants could be biologically indistinguishable from the first, in which case we are in the replication scenario; or the variants could carry slightly different implications, in which case we are instead facing an extension rather than a replication. The conceptual framework and the practical method we propose apply to both cases: whether we are dealing with a replication or an extension, we can study the agreement between effects, although, depending on the case, we are asking different questions.

\section{Generative Model and Replicability Metrics}\label{sec:model}
In the preceding sections, we made several distinctions: between evidence that the effect is above a threshold and evidence for its replicability, between replication within a target population and agreement across studies that derive from different target populations, and between replicability metrics computed in the empirical space and replicability metrics computed in the continuous parameter space. We now make these concrete by introducing a hierarchical model from which each replicability question can be read.

In what follows we use \emph{agreement} and \emph{consistency} in specific technical senses: agreement for concordance between the study-specific effects, consistency for concordance of the study-specific effects with the generative effect.

\subsection{The Model}\label{sec:generativemodel}
Consider a set of $S$ studies, with $s = 1,\dots,S$ indexing the study and $i = 1,\dots,n_s$ indexing the individuals within it. We model the data through a hierarchical regression. For a continuous outcome $Y_{si}$ and a predictor of interest $X_{si}$, the model is
\begin{equation}\label{eq:model}
  Y_{si} \mid \alpha_s, \beta_s, \sigma_s \;\sim\;
\mathcal{N}\!\left(\alpha_s + \beta_s X_{si},\, \sigma_s^2\right),
\end{equation} 
where $\alpha_s$ is the study-specific intercept, $\beta_s$ the study-specific effect of interest, and $\sigma_s^2$ the study-specific residual variance.

The study-specific parameters are drawn from super-population-level distributions:
\begin{align}
  \beta_s \mid \beta, \tau_\beta &\sim \text{Student-}t(\nu, \beta, \tau_\beta), \\
  \alpha_s \mid \alpha, \tau_\alpha &\sim \text{Student-}t_{[0, \infty)}(\nu, \alpha, \tau_\alpha), \\
  \sigma_s \mid \sigma, \tau_\sigma &\sim \text{Student-}t_{[0, \infty)}(\nu, \sigma, \tau_\sigma),
\end{align}
where all three Student-$t$ distributions share $\nu = 3$ degrees of freedom, to accommodate heterogeneity between studies.
The \emph{generative} model specifies the full distribution that generates collections of studies, and thus yields a predictive distribution for future studies. Within it, we call $\text{Student-}t(\nu, \beta, \tau_\beta)$ the \emph{generative distribution} of the effects and its location $\beta$ the \emph{generative effect}: it is the shared generative effect from whose distribution the study-specific effects $\beta_s$ are drawn. The parameters $\beta$, $\alpha$ and $\sigma$ and $\tau_\beta$, $\tau_\alpha$ and $\tau_\sigma$ are, respectively, the locations and scales of the corresponding $\text{Student-}t$ distributions. All of these parameters are treated as random, each with its own hyperprior: we write $\mu_\beta, \mu_\alpha, \mu_\sigma$ and $\eta_\beta, \eta_\alpha, \eta_\sigma$ for the location and scale of the hyperpriors on $\beta, \alpha, \sigma$, and $\mu_{\tau_\beta}, \mu_{\tau_\alpha}, \mu_{\tau_\sigma}$ and $\eta_{\tau_\beta}, \eta_{\tau_\alpha}, \eta_{\tau_\sigma}$, for those on the heterogeneity scales. Both the distributional choices and the twelve hyperparameters, elicited by an empirical Bayes procedure, are explained in Section~\ref{sec:empirical}.

That the heterogeneity scales are governed by hyperprior locations $\mu_{\tau_\beta}, \mu_{\tau_\alpha}, \mu_{\tau_\sigma}$ is what lets the model span the range from independent to fully pooled analyses as a limiting behavior. Treating the studies as entirely independent, without shared information between them, corresponds to the heterogeneity priors concentrating at ever higher locations, $\mu_{\tau_\beta}, \mu_{\tau_\alpha}, \mu_{\tau_\sigma} \to \infty$; the complete pooling corresponds to the opposite limit, the locations approaching zero. Therefore, moving from independence to complete pooling is a matter of where the heterogeneity priors are placed, not a change of model. The exact independence limit cannot be realized as a proper, computable model: at the limit the prior on each $\beta_s$ is improper and $\beta$ is unidentified. We therefore propose to approximate the limit by fixing each heterogeneity scale $\tau_\beta,\tau_\alpha,\tau_\sigma$ to the 99.9th percentile of its respective prior distribution, a finite value that decouples the studies while keeping the model proper.

This continuum also gives a precise form to the distinction between replication and extension introduced in Section~\ref{sec:glossary}.
The limit $\mu_{\tau_\beta}, \mu_{\tau_\alpha}, \mu_{\tau_\sigma} \to 0$ corresponds to the replication scenario: the studies derive from the same target population. If the intercept and residual scale differ between studies, the comparison becomes an extension, whether the study-specific effects still derive from the same target population ($\mu_{\tau_\beta} \to 0$) or are themselves allowed to differ ($\mu_{\tau_\beta} > 0$), as defined in Section~\ref{sec:replication_extension}. The prior distributions on the heterogeneity parameters encode the researcher's belief about whether the comparison is a replication or an extension, through the
amount of prior mass placed near zero: the more prior mass concentrated near zero, the stronger the belief in a replication. 

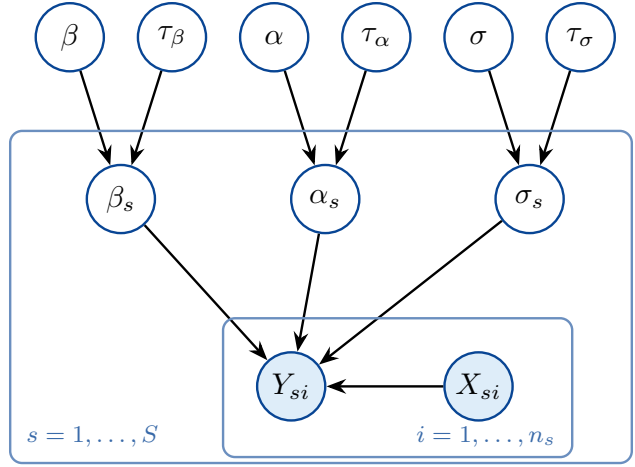
\begin{figure}[ht]
    \centering
    \definecolor{DeepBlue}{RGB}{8, 69, 148}
    \definecolor{Highlight}{RGB}{200, 225, 245}
    \resizebox{0.5\textwidth}{!}{%
    \begin{tikzpicture}[
        >=Stealth, thick,
        every node/.style={font=\small},
        latent/.style={circle, draw=DeepBlue, fill=white, minimum size=0.8cm, inner sep=0pt},
        obs/.style={circle, draw=DeepBlue, fill=Highlight!60, minimum size=0.8cm, inner sep=0pt}
    ]
    \node[latent] (theta) at (-2.4, 2.5) {$\beta$};
    \node[latent] (taut)  at (-1.2, 2.5) {$\tau_\beta$};
    \node[latent] (alpha) at ( 0.0, 2.5) {$\alpha$};
    \node[latent] (taua)  at ( 1.2, 2.5) {$\tau_\alpha$};
    \node[latent] (sigma) at ( 2.4, 2.5) {$\sigma$};
    \node[latent] (taus)  at ( 3.6, 2.5) {$\tau_\sigma$};
    \node[latent] (tr)    at (-1.8, 0.6) {$\beta_s$};
    \node[latent] (ar)    at ( 0.6, 0.6) {$\alpha_s$};
    \node[latent] (sr)    at ( 3.0, 0.6) {$\sigma_s$};
    \node[obs]    (Y)     at ( 0.2,-1.6) {$Y_{si}$};
    \node[obs]    (X)     at ( 2.4,-1.6) {$X_{si}$};
    \draw[->] (theta) -- (tr); \draw[->] (taut) -- (tr);
    \draw[->] (alpha) -- (ar); \draw[->] (taua) -- (ar);
    \draw[->] (sigma) -- (sr); \draw[->] (taus) -- (sr);
    \draw[->] (tr) -- (Y); \draw[->] (ar) -- (Y); \draw[->] (sr) -- (Y);
    \draw[->] (X) -- (Y);
    \draw[rounded corners=4pt, DeepBlue!60] (-3.1,-2.5) rectangle (4.2, 1.4);
    \node[anchor=south west, font=\scriptsize\color{DeepBlue!80}] at (-3.05,-2.45) {$s=1,\dots,S$};
    \draw[rounded corners=4pt, DeepBlue!60] (-0.6,-2.43) rectangle (3.5,-0.8);
    \node[anchor=south east, font=\scriptsize\color{DeepBlue!80}] at (3.45,-2.45) {$i=1,\dots,n_s$};
    \end{tikzpicture}%
    }
    \caption{Directed Acyclic Graph summarizing the notation and conditional independence assumptions of the generative model.}
    \label{fig:plate_notation}
\end{figure}

Figure~\ref{fig:plate_notation} shows the directed acyclic graph that summarizes the notation and conditional independence assumptions of the generative model. The shaded nodes are the observed data, while the unshaded nodes are the study-specific parameters and their respective generative parameters.

\subsection{Region of Practical Relevance}
Let $t = 1,\dots,T$ index the posterior draws obtained by Markov chain Monte Carlo. For each study $s$, the posterior provides draws $\beta_s^{(t)}$ of the study-specific effect, and draws $\beta^{(t)}$ of the generative effect. The replicability metrics developed below are all posterior probabilities of events expressed through these draws.

\begin{definition}[Region of practical relevance]\label{def:ropr}
Let $\varepsilon > 0$ be a threshold such that any effect $\theta$ with $|\theta| \le \varepsilon$ is taken to be of practical equivalence to the null. The interval $[-\varepsilon,\varepsilon]$ is the region of practical equivalence to the null \citep{kruschke2018rejecting}; its complement,
\begin{equation}
  \mathcal{B} = \{\, \theta \in \mathbb{R} : |\theta| > \varepsilon \,\},
\end{equation}
is the region of practical relevance. An effect is said to be of practical relevance when it lies in $\mathcal{B}$.
\end{definition}
A defensible threshold $\varepsilon$ is typically based on domain-specific knowledge or on generic conventions \citep{kruschke2018rejecting}, and is best elicited from considerations analogous to the smallest effect size of interest in equivalence testing \citep{lakens2018equivalence, lakens2017equivalence} or the minimal clinically important difference in medicine \citep{jaeschke1989measurement}.

Throughout, for a generic effect $\theta$ we say that $\theta$ is \emph{above} $\varepsilon$ when $\theta>\varepsilon$ and \emph{below} $-\varepsilon$ when $\theta<-\varepsilon$; $\theta$ is \emph{of practical relevance} when $|\theta|>\varepsilon$, and \emph{of practical equivalence to the null} when $|\theta|\le\varepsilon$. The same terminology applies both to the generative effect $\beta$ and to the study-specific effects $\beta_s$.

\subsection{The Generative Effect}

The simplest metric concerns the generative effect alone. It asks whether the generative effect is of practical relevance.

\begin{definition}[Generative effect relevance, $P_\beta$]\label{def:pbeta}
The posterior probability that the generative effect lies in the region of practical relevance is
\begin{equation}
  P_\beta = P(\beta \in \mathcal{B} \mid \text{data})
          = \frac{1}{T}\sum_{t=1}^{T} \mathbf{1}\big(|\beta^{(t)}| > \varepsilon\big).
\end{equation}
\end{definition}

When replicability is assessed across several studies ($S \ge 2$), requiring unanimity among all of them is a particularly strict criterion. To allow for partial agreement, we record, at each posterior draw $t$, how many studies fall in each region:
\begin{equation}
\begin{aligned}
N_{pos}^{(t)}  &= \sum_{s=1}^{S} \mathbf{1}\big(\beta_s^{(t)} > \varepsilon\big); \\
N_{neg}^{(t)}  &= \sum_{s=1}^{S} \mathbf{1}\big(\beta_s^{(t)} < -\varepsilon\big); \\
N_{null}^{(t)} &= \sum_{s=1}^{S} \mathbf{1}\big(|\beta_s^{(t)}| \le \varepsilon\big).
\end{aligned}
\end{equation}
The metrics below are built on a \emph{consensus level} $k \in \{2,\dots,S\}$: we require at least $k$ of the $S$ study-specific effects to agree, with $k = S$ recovering unanimity and $k < S$ admitting partial agreement, leaving the remaining $S-k$ effects free.

\subsection{Agreement among Studies}\label{sec:nogenerativemetrics}
The metrics defined in this section measure the agreement, regarding the effect of interest, between studies. These metrics are available whenever a collection of study-specific effects is compared, whether or not a generative effect is entertained, which is why they also describe the limiting case in which the studies are treated as independent.

\begin{definition}[Overall agreement, $P_{\text{overall},k}$]\label{def:overall_agreement}
The posterior probability that at least $k=2,\dots,S$ study-specific effects $\beta_s$ are jointly above $\varepsilon$, jointly below $-\varepsilon$, or jointly of practical equivalence to the null, is
\begin{equation}
P_{\text{overall},k}
=
\frac{1}{T}\sum_{t=1}^{T}
\mathbf{1}\Big(
N_{pos}^{(t)} \ge k \;\vee\; N_{neg}^{(t)} \ge k \;\vee\; N_{null}^{(t)} \ge k
\Big).
\end{equation}
\end{definition}

When the consensus threshold $k$ exceeds half of the studies, $k > S/2$, the three conditions in Definition~\ref{def:overall_agreement} are mutually exclusive, and the overall agreement decomposes into three separate metrics:

\begin{equation}
P_{\text{overall},k} = P_{\text{pos},k} + P_{\text{neg},k} + P_{\text{null},k}.
\end{equation}
The exclusivity follows because $N_{pos}^{(t)} \ge k$ and $N_{neg}^{(t)} \ge k$ together would require $N_{pos}^{(t)} + N_{neg}^{(t)} \ge 2k > S$, which is impossible since $N_{pos}^{(t)} + N_{neg}^{(t)} + N_{null}^{(t)} = S$. For even $S$ this excludes the exact-half value $k = S/2$: a draw with $S/2$ effects above $\varepsilon$ and $S/2$ below $-\varepsilon$ would count in both $P_{\text{pos},k}$ and $P_{\text{neg},k}$, so the smallest admissible majority is $k = S/2 + 1$.

The decomposition is useful when the scientific question is directional, for instance when one asks specifically whether the treatment increases the level of the outcome variable.

\begin{definition}[Positive agreement, $P_{\text{pos},k}$]
The posterior probability that at least $k=2,\dots,S$ study-specific effects $\beta_s$ agree that the effect is above $\varepsilon$ is
\begin{equation}
P_{\text{pos},k} = \frac{1}{T}\sum_{t=1}^{T} \mathbf{1}\big(N_{pos}^{(t)} \ge k\big).
\end{equation}
\end{definition}

\begin{definition}[Negative agreement, $P_{\text{neg},k}$]
The posterior probability that at least $k=2,\dots,S$ study-specific effects $\beta_s$ agree that the effect is below $-\varepsilon$ is
\begin{equation}
P_{\text{neg},k} = \frac{1}{T}\sum_{t=1}^{T} \mathbf{1}\big(N_{neg}^{(t)} \ge k\big).
\end{equation}
\end{definition}

\begin{definition}[Null agreement, $P_{\text{null},k}$]
The posterior probability that at least $k=2,\dots,S$ study-specific effects $\beta_s$ agree that the effect is within $[-\varepsilon, \varepsilon]$, is
\begin{equation}
P_{\text{null},k} = \frac{1}{T}\sum_{t=1}^{T} \mathbf{1}\big(N_{null}^{(t)} \ge k\big).
\end{equation}
\end{definition}

\begin{definition}[Non-null agreement, $P_{\text{non-null},k}$]
The posterior probability that at least $k=2,\dots,S$ study-specific effects $\beta_s$ agree that the effect is of practical relevance and of the same sign,
i.e.\ jointly above $\varepsilon$ or jointly below $-\varepsilon$,
excluding agreement on an effect of practical equivalence to the null, is
\begin{equation}
P_{\text{non-null},k}
=
\frac{1}{T}\sum_{t=1}^{T}
\mathbf{1}\big(
N_{pos}^{(t)} \ge k \;\vee\; N_{neg}^{(t)} \ge k
\big).
\end{equation}
\end{definition}

\begin{definition}[Conditional probability, $P_{\text{cond}\mid O,m}$]
Let study $O \in \{1,\dots,S\}$ be a reference study, and let \begin{equation}
N_{pos}^{(-O),(t)} = \sum_{s \neq O} \mathbf{1}\big(\beta_s^{(t)} > \varepsilon\big), \
N_{neg}^{(-O),(t)} = \sum_{s \neq O} \mathbf{1}\big(\beta_s^{(t)} < -\varepsilon\big)
\end{equation}
count, at each draw, how many study-specific effects $\beta_s$, other than $\beta_O$, are above $\varepsilon$ and how many below $-\varepsilon$. The conditional posterior probability that the effect of practical relevance found in the reference study $O$ recurs, in the same direction and in at least $m = 1,\dots,S-1$ of the other studies, is
\begin{equation}
P_{\text{cond}\mid O, m}
=
\frac{
\sum_{t=1}^{T} \mathbf{1}\!\left(
\begin{aligned}
&\big(\beta_O^{(t)} > \varepsilon \;\wedge\; N_{pos}^{(-O),(t)} \ge m \big) \;\vee\; \\
&\big(\beta_O^{(t)} < -\varepsilon \;\wedge\; N_{neg}^{(-O),(t)} \ge m \big)
\end{aligned}
\right)
}{
\sum_{t=1}^{T} \mathbf{1}\big( |\beta_O^{(t)}| > \varepsilon \big)
}.
\end{equation}
\end{definition}
For the conditional probability the threshold $m$ refers to the number of other studies required to corroborate the discovery.

Figure~\ref{fig:indep_probs} gives geometric intuition for these metrics in the two-study case $(S = 2,\ k = 2)$, showing the subspaces of the parameter space that satisfy each condition.

\definecolor{DeepBlue}{RGB}{44, 62, 80}
\definecolor{Highlight}{RGB}{200, 225, 245}

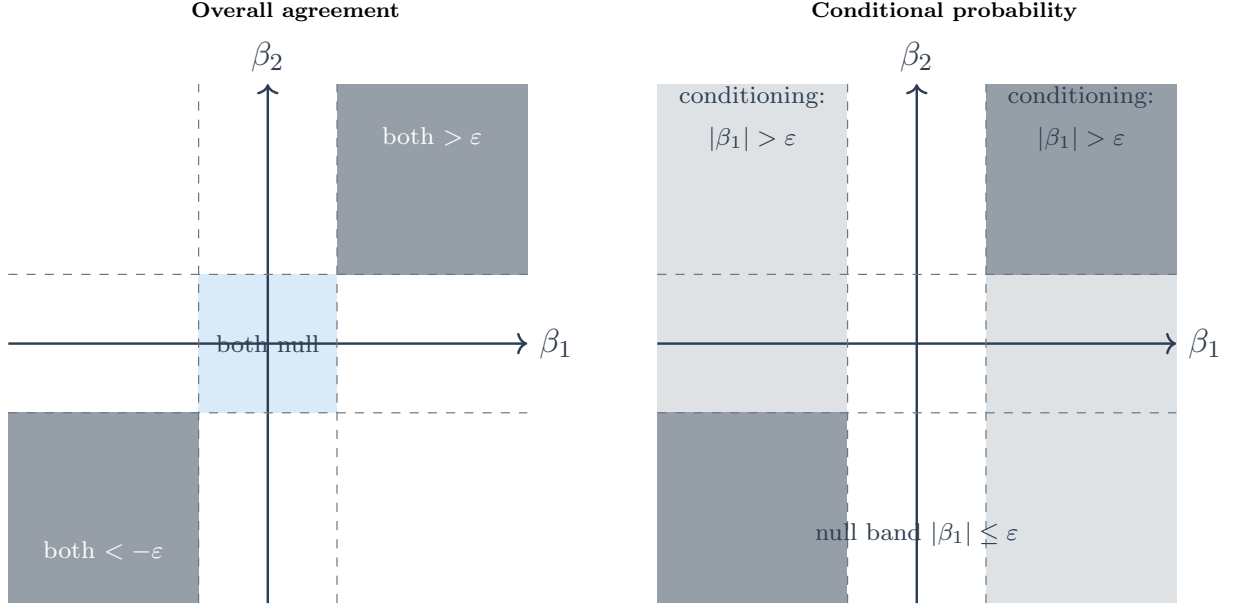
\begin{figure}[ht]
    \centering
    \begin{minipage}{0.48\linewidth}
        \centering
        \textbf{\scriptsize Overall agreement}\par\vspace{0.1cm}
        \resizebox{\linewidth}{!}{
        \begin{tikzpicture}
            \fill[DeepBlue!50] (0.8,0.8) rectangle (3,3);
            \fill[DeepBlue!50] (-3,-3) rectangle (-0.8,-0.8);
            \fill[Highlight!70] (-0.8,-0.8) rectangle (0.8,0.8);
            \draw[->, thick, DeepBlue] (-3,0) -- (3,0) node[right] {$\beta_1$};
            \draw[->, thick, DeepBlue] (0,-3) -- (0,3) node[above] {$\beta_2$};
            \draw[dashed, DeepBlue!70] (-3,0.8) -- (3,0.8);
            \draw[dashed, DeepBlue!70] (-3,-0.8) -- (3,-0.8);
            \draw[dashed, DeepBlue!70] (0.8,-3) -- (0.8,3);
            \draw[dashed, DeepBlue!70] (-0.8,-3) -- (-0.8,3);
            \node[text=white] at (1.9,2.4) {\scriptsize both $>\varepsilon$};
            \node[text=white] at (-1.9,-2.4) {\scriptsize both $<-\varepsilon$};
            \node[text=DeepBlue] at (0,0) {\scriptsize both null};
        \end{tikzpicture}
        }
    \end{minipage}%
    \hfill
    \begin{minipage}{0.48\linewidth}
        \centering
        \textbf{\scriptsize Conditional probability}\par\vspace{0.1cm}
        \resizebox{\linewidth}{!}{
        \begin{tikzpicture}
            \fill[DeepBlue!15] (-3,3) rectangle (-0.8,-3);
            \fill[DeepBlue!15] (0.8,-3) rectangle (3,3);
            \fill[DeepBlue!50] (0.8,0.8) rectangle (3,3);
            \fill[DeepBlue!50] (-3,-3) rectangle (-0.8,-0.8);
            \fill[white] (-0.8,-0.8) rectangle (0.8,0.8);  
            \fill[white] (-0.8,0.8) rectangle (0.8,3);
            \fill[white] (-0.8,-3) rectangle (0.8,-0.8);
            \draw[->, thick, DeepBlue] (-3,0) -- (3,0) node[right] {$\beta_1$};
            \draw[->, thick, DeepBlue] (0,-3) -- (0,3) node[above] {$\beta_2$};
            \draw[dashed, DeepBlue!70] (-3,0.8) -- (3,0.8);
            \draw[dashed, DeepBlue!70] (-3,-0.8) -- (3,-0.8);
            \draw[dashed, DeepBlue!70] (0.8,-3) -- (0.8,3);
            \draw[dashed, DeepBlue!70] (-0.8,-3) -- (-0.8,3);
            \node[text=DeepBlue, align=center] at (-1.9,2.6) {\scriptsize conditioning:\\ \scriptsize $|\beta_1|>\varepsilon$};
            \node[text=DeepBlue, align=center] at (1.9,2.6)  {\scriptsize conditioning:\\ \scriptsize $|\beta_1|>\varepsilon$};
            \node[text=DeepBlue] at (0,-2.2) {\scriptsize null band $|\beta_1|\le\varepsilon$};
        \end{tikzpicture}
        }
    \end{minipage}
    \caption{Subsets of the parameter space defining agreement in the
two-study case $(S = 2)$. Left, the agreement regions: in the top-right square
both effects are above $+\varepsilon$ (the corresponding probability is $P_{\text{pos},2}$); in the bottom-left square, both effects are below $-\varepsilon$ (the corresponding probability is $P_{\text{neg},2}$); in the central square, both effects are of practical equivalence to the null (the corresponding probability gives $P_{\text{null},2}$). The
sum of these probabilities is the overall agreement $P_{\text{overall},2}$. The sum of the first two is the non-null agreement $P_{\text{non-null},2}$. Right, relevant subsets for the conditional probability $P_{\text{cond}\mid O=1, m=1}$: the vertical bands denote the conditioning event $|\beta_1| > \varepsilon$, their darker subsets the replication of that effect in the same direction in the second study. The white central band, where $|\beta_1| \le \varepsilon$, falls outside the conditioning event.}
\label{fig:indep_probs}
\end{figure}

\subsection{Consistency with the Generative Effect}\label{sec:generativemetrics}
The second family of metrics draws on the generative level. Using the full hierarchical structure, we can now ask a further question: what is the probability that the study-specific effects are consistent with the generative effect? Since these metrics require a generative effect $\beta$ to be consistent with, there is no counterpart when the studies are treated as independent.

\begin{definition}[Generative overall consistency, \\ $P_{\text{gen,overall},k}$]
The posterior probability that at least $k=2,\dots,S$ study-specific effects $\beta_s$ and the generative effect $\beta$ are jointly above $\varepsilon$, jointly below $-\varepsilon$ or jointly of practical equivalence to the null, is
\begin{equation}
P_{\text{gen,overall},k} = \frac{1}{T} \sum_{t=1}^{T} \mathbf{1}\!\left(
\begin{aligned}
&(N_{pos}^{(t)} \ge k \wedge \beta^{(t)} > \varepsilon) \;\vee\; \\
&(N_{neg}^{(t)} \ge k \wedge \beta^{(t)} < -\varepsilon) \;\vee\; \\
&(N_{null}^{(t)} \ge k \wedge |\beta^{(t)}| \le \varepsilon)
\end{aligned}
\right).
\end{equation}
\label{def:gen_overall}
\end{definition}

As with the agreement metrics, the generative overall consistency (Definition~\ref{def:gen_overall}) decomposes into its three components,
\begin{equation}
P_{\text{gen,overall},k} = P_{\text{gen,pos},k} + P_{\text{gen,neg},k} + P_{\text{gen,null},k}.
\end{equation}
Here the three components are mutually exclusive for any $k$, since they are distinguished by the sign of the generative effect $\beta$, which is above $\varepsilon$, below $-\varepsilon$, or within $[ -\varepsilon,\varepsilon]$ in each draw but never more than one of these at once.

\begin{definition}[Generative positive consistency, \\ $P_{\text{gen,pos},k}$]
The posterior probability that at least $k=2,\dots,S$ study-specific effects $\beta_s$ and the generative effect $\beta$ agree that the effect is above $\varepsilon$ is
\begin{equation}
P_{\text{gen,pos},k} =
\frac{1}{T} \sum_{t=1}^{T}
\mathbf{1}\big(
N_{pos}^{(t)} \ge k \;\wedge\; \beta^{(t)} > \varepsilon
\big).
\end{equation}
\end{definition}

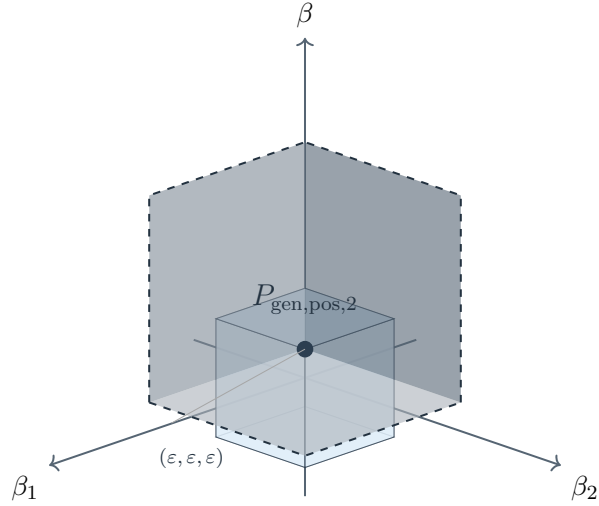
\begin{figure}[ht]
\centering
\textbf{\scriptsize Generative positive consistency}
\par\vspace{0.6cm}
\tdplotsetmaincoords{70}{135}
\resizebox{0.5\textwidth}{!}{
    \begin{tikzpicture}[tdplot_main_coords, scale=2.2]
    \def\e{0.4}
    \def\L{1.8}
    \draw[->, thick, DeepBlue!80] (-1,0,0) -- (\L+0.5,0,0) node[anchor=north east, text=black] {$\beta_1$};
    \draw[->, thick, DeepBlue!80] (0,-1,0) -- (0,\L+0.5,0) node[anchor=north west, text=black] {$\beta_2$};
    \draw[->, thick, DeepBlue!80] (0,0,-0.8) -- (0,0,\L+0.5) node[anchor=south, text=black] {$\beta$};
    \filldraw[Highlight!30, opacity=0.55, draw=DeepBlue!70] (-\e,-\e,-\e) -- (-\e,\e,-\e) -- (-\e,\e,\e) -- (-\e,-\e,\e) -- cycle;
    \filldraw[Highlight!30, opacity=0.55, draw=DeepBlue!70] (-\e,-\e,-\e) -- (\e,-\e,-\e) -- (\e,-\e,\e) -- (-\e,-\e,\e) -- cycle;
    \filldraw[Highlight!30, opacity=0.55, draw=DeepBlue!70] (-\e,-\e,-\e) -- (\e,-\e,-\e) -- (\e,\e,-\e) -- (-\e,\e,-\e) -- cycle;
    \filldraw[Highlight!60, opacity=0.80, draw=DeepBlue] (\e,-\e,-\e) -- (\e,\e,-\e) -- (\e,\e,\e) -- (\e,-\e,\e) -- cycle;
    \filldraw[Highlight!60, opacity=0.80, draw=DeepBlue] (-\e,\e,-\e) -- (\e,\e,-\e) -- (\e,\e,\e) -- (-\e,\e,\e) -- cycle;
    \filldraw[Highlight!75, opacity=0.85, draw=DeepBlue] (-\e,-\e,\e) -- (\e,-\e,\e) -- (\e,\e,\e) -- (-\e,\e,\e) -- cycle;
    \fill[DeepBlue!40, opacity=0.6] (\e,\e,\e) -- (\L,\e,\e) -- (\L,\L,\e) -- (\e,\L,\e) -- cycle;
    \fill[DeepBlue!60, opacity=0.6] (\e,\e,\e) -- (\L,\e,\e) -- (\L,\e,\L) -- (\e,\e,\L) -- cycle;
    \fill[DeepBlue!80, opacity=0.6] (\e,\e,\e) -- (\e,\L,\e) -- (\e,\L,\L) -- (\e,\e,\L) -- cycle;
    \draw[dashed, DeepBlue!80!black, thick] (\L,\e,\e) -- (\L,\L,\e) -- (\e,\L,\e);
    \draw[dashed, DeepBlue!80!black, thick] (\L,\e,\e) -- (\L,\e,\L) -- (\e,\e,\L);
    \draw[dashed, DeepBlue!80!black, thick] (\e,\L,\e) -- (\e,\L,\L) -- (\e,\e,\L);
    \node[text=DeepBlue!80!black, font=\large] at (\e+0.7, \e+0.7, \e+0.7) {${P_{\text{gen,pos},2}}$};
    \fill[DeepBlue] (\e,\e,\e) circle (1.5pt);
    \draw[thin, gray!70] (\e,\e,\e) -- (\e+0.6,\e-0.6,\e-0.5);
    \node[font=\scriptsize, text=DeepBlue!80!black, anchor=north west] at (\e+0.7,\e-0.7,\e-0.6) {$(\varepsilon,\varepsilon,\varepsilon)$};
    \end{tikzpicture}
}
\caption{The subset of the parameter space defining generative positive consistency $P_{\text{gen,pos},2}$ in the two-study case $(S = 2)$. The central pale cube is the region of practical equivalence to the null, $|\,\cdot\,| \le \varepsilon$ on every axis. The unbounded dark region is the subset in which the study-specific effects $\beta_1, \beta_2$ and the generative effect $\beta$ are simultaneously above the threshold $\varepsilon$.}
\label{fig:3d_region}
\end{figure}

Figure~\ref{fig:3d_region} gives geometric intuition for the positive generative consistency $P_{\text{gen,pos},2}$ in the two-study case $(S = 2,\ k = 2)$.

\begin{definition}[Generative negative consistency, \\ $P_{\text{gen,neg},k}$]
The posterior probability that at least $k=2,\dots,S$ study-specific effects $\beta_s$ and the generative effect $\beta$ agree that the effect is below $-\varepsilon$ is
\begin{equation}
P_{\text{gen,neg},k} =
\frac{1}{T} \sum_{t=1}^{T}
\mathbf{1}\big(
N_{neg}^{(t)} \ge k \;\wedge\; \beta^{(t)} < -\varepsilon
\big).
\end{equation}
\end{definition}

\begin{definition}[Generative null consistency, \\ $P_{\text{gen,null},k}$]
The posterior probability that at least $k=2,\dots,S$ study-specific effects $\beta_s$ and the generative effect $\beta$ agree that the effect is within [$-\varepsilon, \varepsilon$], is
\begin{equation}
P_{\text{gen,null},k} =
\frac{1}{T} \sum_{t=1}^{T}
\mathbf{1}\big(
N_{null}^{(t)} \ge k \;\wedge\; |\beta^{(t)}| \le \varepsilon
\big).
\end{equation}
\end{definition}

\begin{definition}[Generative non-null consistency, \\ $P_{\text{gen,non-null},k}$]
The posterior probability that at least $k=2,\dots,S$ study-specific effects $\beta_s$ and the generative effect $\beta$ agree that the effect is of practical relevance and of the same sign, i.e.\ jointly above $\varepsilon$ or jointly below $-\varepsilon$, excluding agreement on an effect of practical equivalence to the null, is
\begin{equation}
P_{\text{gen,non-null},k} = \frac{1}{T} \sum_{t=1}^{T} \mathbf{1}\!\left(
\begin{aligned}
&(N_{pos}^{(t)} \ge k \wedge \beta^{(t)} > \varepsilon) \;\vee\; \\
&(N_{neg}^{(t)} \ge k \wedge \beta^{(t)} < -\varepsilon)
\end{aligned}
\right).
\end{equation}
\end{definition}
The generative consistency metrics admit a conditional counterpart, obtained by restricting attention to the draws in which the generative effect takes a given sign.

\begin{definition}[Conditional generative positive consistency, $P_{\text{cpos},k}$]
Conditional on the generative effect $\beta$ being above $\varepsilon$, the probability that it is supported by at least $k=2,\dots,S$ study-specific effects $\beta_s$, is
\begin{equation}
P_{\text{cpos},k}
=
\frac{
\sum_{t=1}^{T}
\mathbf{1}\big( N_{pos}^{(t)} \ge k \;\wedge\; \beta^{(t)} > \varepsilon \big)
}{
\sum_{t=1}^{T}
\mathbf{1}\big(\beta^{(t)} > \varepsilon\big)
}.
\end{equation}
\end{definition}

\begin{definition}[Conditional generative negative consistency, $P_{\text{cneg},k}$]
Conditional on the generative effect $\beta$ being below $-\varepsilon$, the probability that it is supported by at least $k=2,\dots,S$ study-specific effects $\beta_s$, is
\begin{equation}
P_{\text{cneg},k}
=
\frac{
\sum_{t=1}^{T}
\mathbf{1}\big( N_{neg}^{(t)} \ge k \;\wedge\; \beta^{(t)} < -\varepsilon \big)
}{
\sum_{t=1}^{T}
\mathbf{1}\big(\beta^{(t)} < -\varepsilon\big)
}.
\end{equation}
\label{def:cneg}
\end{definition}

\begin{definition}[Conditional generative null consistency, $P_{\text{cnull},k}$]
Conditional on the generative effect $\beta$ being of practical equivalence to the null, the probability that it is supported by at least $k=2,\dots,S$ study-specific effects $\beta_s$, is
\begin{equation}
P_{\text{cnull},k}
=
\frac{
\sum_{t=1}^{T}
\mathbf{1}\big( N_{null}^{(t)} \ge k \;\wedge\; |\beta^{(t)}| \le \varepsilon \big)
}{
\sum_{t=1}^{T}
\mathbf{1}\big(|\beta^{(t)}| \le \varepsilon\big)
}.
\end{equation}
\end{definition}

A generative effect $\beta$ near zero can arise in two ways: when both the generative effect $\beta$ and every study-specific effect $\beta_s$ are of practical equivalence to the null, or when the studies carry strong but opposing effects that average to near zero, so that a meta-analysis infers absence while an effect of practical relevance is present in some studies. The joint probability $P_{\text{gen,null},k}$, which requires a generative effect of practical equivalence to the null, is high only if both the generative effect $\beta$ and the study-specific effects $\beta_s$ are of practical equivalence to the null. On its own, however, a near-zero joint is ambiguous: it is small both when the generative effect $\beta$ is of practical equivalence to the null and at the same time few studies are of practical equivalence to the null (so $N_{null}^{(t)}$ is small), and when the generative effect and the study-specific effects are all of practical relevance.

The conditional $P_{\text{cnull},k}$ helps distinguish the two ways in which a near-zero generative effect can arise. Conditioning on $\beta$ being of practical equivalence to the null, it is high when the studies that produced it are themselves of practical equivalence to the null, and collapses when the study-specific effects point in opposing directions, outside $[-\varepsilon,\varepsilon]$, even though the generative effect lands near zero. The same near-zero joint thus carries opposite meanings according to the conditional beside it. The marginal probability $P(|\beta|\le\varepsilon)$ plays the supporting role of saying when the conditional can be trusted: when it is large, $P_{\text{cnull},k}$ rests on many draws and is reliable; when it is small, the case of a truly present effect, where $\beta$ is rarely of practical equivalence to the null, it becomes an unstable near-$0/0$ ratio, and one should rely on the directional metrics instead.

Therefore, no single metric is meant to be read in isolation; it is the overall picture that is informative about the phenomenon of interest. Study-specific effects may agree with one another yet not with the generative one, agree on an effect of practical relevance that the generative effect does not share, or concur on an effect of practical equivalence to the null that the generative effect contradicts, and each such configuration produces a distinct pattern across the agreement, generative consistency, and conditional metrics. We return to these patterns on the empirical application in Section~\ref{sec:empirical}, where the metrics are computed and read together.

\section{A Unified Model for Replicability Questions}
Having defined the metrics, we now show how the single model can answer a range of replicability questions through them.

\subsection{Two Aspects of the Unified Framework}\label{sec:unified}
Our framework is unified in two senses: we propose one model in place of a fragmented collection of methods, and one family of metrics in place of a different summary for each question.

The first sense is that the single model specified in Section~\ref{sec:generativemodel} answers some of the scientific questions that the literature addresses with separate and often mutually inconsistent tools, as explained in Section~\ref{sec:sota}. The scientific questions we are able to address can be summarized in the following way:

\begin{enumerate}
    \item \textit{Agreement between the study-specific effects.} Treating the study-specific effects $\beta_1,\dots,\beta_S$ as the objects of interest, one asks how the individual studies agree with one another. This concept of replicability compares the study-specific effects only with one another, recovering the notion as it is traditionally posed for a collection of independent studies. We refer to this notion as \emph{horizontal} replicability.

    \item \textit{Consistency of the study-specific effects with the generative effect.}
    Analyzing $\beta_s$ and $\beta$ together, one asks whether the study-specific effects $\beta_s$ are consistent with the generative effect $\beta$. We refer to this notion as \emph{vertical} replicability.

    \item \textit{Agreement between the study-specific effect predicted from the body of evidence and the effect observed in the left-out study.}
    Writing $\mathcal{D}$ for the data of all $S$ studies and $\mathcal{D}_{-\ell}$ for the data of all studies except a left-out study $\ell$, we fit the generative model to $\mathcal{D}_{-\ell}$ to predict the effect $\beta_\ell^{\mathrm{pred}}$ of the left-out study, and we ask whether the predicted effect $\beta_\ell^{\mathrm{pred}}$, obtained from the existing evidence $\mathcal{D}_{-\ell}$, agrees with the effect $\beta_\ell$ observed in the left-out study.

    \item \textit{Inference on the generative effect.} 
    Treating the generative effect $\beta$ as the object of interest, one asks what the super-population of studies implies about the generative effect.
\end{enumerate}

As shown in Table~\ref{tab:questions}, each question becomes a functional of the same model, depending only on which parameters one interrogates. The novelty lies here: not in the hierarchical model, nor in any single question it answers, but in posing all of them within one model and in using the continuous parameter space to quantify different and complementary nuances of replicability. 

The metrics corresponding to the predictive question (the third one in Table~\ref{tab:questions}) are computed with the same definitions as those for the first question, introduced in Section~\ref{sec:nogenerativemetrics}: overall, positive, negative, null and non-null agreement. The only difference is that for the predictive question the two compared distributions are not the posteriors of two observed effects, but the posterior of one observed effect and the predictive distribution of the effect that belongs to the left-out study $\ell$. We mark the latter with the subscript $(\cdot)_{\ell}$ solely to distinguish these metrics from those computed between pairs of observed studies. How this predictive distribution is constructed is detailed in Section~\ref{sec:computational}.

Therefore, there is no need to introduce different metrics for each question. This identity of definitions is an example of the second sense in which the framework is unified: the four questions are answered not only by one model but by one family of metrics.

\begin{table}[!ht]
\centering
\caption{The four replicability questions answered by the model, the parameters
each interrogates, and the metrics defined in this section.}
\label{tab:questions}
\small
\begin{tabular}{p{3.2cm}ll}
\toprule
Question & Parameters & Metrics \\
\midrule
Study-specific effects
 & $\beta_1,\dots,\beta_S$
 & $P_{\text{overall},k}$, \\
 & & $P_{\text{pos},k}$, \\
 & & $P_{\text{neg},k}$, \\
 & & $P_{\text{null},k}$, \\
 & & $P_{\text{non-null},k}$, \\
 & & $P_{\text{cond}\mid O,m}$. \\[3pt]
Study-specific effects vs.\ generative effect
 & $\beta_1,\dots,\beta_S,\ \beta$
 & $P_{\text{gen,overall},k}$, \\
 & & $P_{\text{gen,pos},k}$, \\
 & & $P_{\text{gen,neg},k}$, \\
 & & $P_{\text{gen,null},k}$, \\
 & & $P_{\text{gen,non-null},k}$, \\
 & & $P_{\text{cpos},k}$, \\
 & & $P_{\text{cneg},k}$, \\
 & & $P_{\text{cnull},k}$. \\[3pt]
Predicted vs.\ observed left-out effect
 & $\beta_\ell^{\mathrm{pred}},\ \beta_\ell$
 & $P_{\text{overall,}\ell}$, \\
 & & $P_{\text{pos,}\ell}$, \\
 & & $P_{\text{neg,}\ell}$, \\
 & & $P_{\text{null,}\ell}$, \\
 & & $P_{\text{non-null,}\ell}$, \\[3pt]
Generative effect
 & $\beta$
 & $P_\beta$. \\
\bottomrule
\end{tabular}
\end{table}

\subsection{Relation to Predictive Approaches}
Our hierarchical model is structurally close to that of \citet{pawel2020probabilistic}, so it is worth analyzing similarities and differences. They study an original--replication pair and assume that each effect estimate is normally distributed around a study-specific effect, with the two study-specific effects drawn from a common generative distribution centered at the generative effect and scattered around it by a between-study variance. We adopt the same generative logic, with study-specific effects $\beta_s$ drawn from a common generative distribution, but with two modeling differences and a conceptual one. 

First, Pawel and Held work with published summaries, specifically the two sample sizes and the original effect estimate, whereas we operate on data at the unit level from both studies and therefore fit the regressions directly. As a consequence, in our model not only the slope but also the intercepts and residual scales are modeled hierarchically.

Second, we treat the between-study heterogeneity as random rather than fixed. Pawel and Held place no prior on the heterogeneity, fixing it at a default value chosen by an elicitation argument and reporting a sensitivity analysis over its magnitude. We instead place priors on $\tau_\beta$, $\tau_\alpha$ and $\tau_\sigma$. In the present application, these priors are elicited empirically (Section~\ref{sec:prior_elicitation}), although the framework is compatible with alternative prior elicitation strategies. As Pawel and Held themselves observe, a full Bayesian treatment of the variance would reflect the uncertainty more accurately \citep{pawel2020probabilistic}.

Finally, there is a conceptual distinction between the two approaches: the role the generative effect plays. The generative effect in \citet{pawel2020probabilistic} establishes that the original and replication estimates are two draws from the same generative distribution, which is what licenses forecasting one from the other. Yet, to obtain the predictive distribution of the replication estimate, the generative effect is integrated out. Their approach builds on earlier methods whose limitations it aims to overcome. Indeed, \citet{patil2016should} obtain essentially the same predictive interval, but treat the studies as sharing a single true effect, and \citet{pawel2020probabilistic} add the hierarchical layer to address the neglect of between-study heterogeneity and the inflation of the original estimate. A similar hierarchical predictive construction had been proposed earlier by \citet{bayarri2002bayesian}. However, both \citet{pawel2020probabilistic} and \citet{bayarri2002bayesian} use the hierarchy to answer the predictive question alone, the agreement of a new study with a body of evidence, whereas we use the same kind of hierarchy to answer all four scientific questions explained in Section~\ref{sec:unified}. Indeed, in our model, the generative effect is not integrated out: it is the object about which some statements are made. The replicability metrics that we propose are posterior probabilities concerning $\beta$ and $\beta_s$. Removing the generative effect would dissolve the questions about the relationship between the study and the generative levels, since there would be no generative effect $\beta$ left to make probability statements about.

As explained so far, the model and the breadth of questions the two approaches answer differ in several respects. One might still expect them to coincide on the predictive question (the third in Table~\ref{tab:questions}), but even there an important difference remains: as throughout our framework, we assess replicability in the continuous parameter space rather than in the empirical one. We obtain the continuous predictive distribution of the effect of a new study and compare it with the posterior distribution of the effect observed in the left-out study, asking whether the two agree in being above $\varepsilon$, below $-\varepsilon$, or within $[-\varepsilon,\varepsilon]$: we thus compare the underlying effects themselves, rather than their observed estimates. By contrast, \citet{pawel2020probabilistic} assess agreement in the empirical space of the estimates: they build a predictive distribution for the replication study's estimate and ask whether the observed estimate is compatible with it.

\subsection{Computational Aspects}\label{sec:computational}
Treating the between-study heterogeneity as random, rather than fixed, has a direct computational consequence. With a normal likelihood and prior and the variance components held fixed, the model is conjugate and the predictive distribution is available in closed form as a normal. Once the heterogeneity is given a prior, however, the predictive must be obtained by integrating over the posterior of that heterogeneity, and the closed form is lost.

Once the model has been fitted to a body of evidence, the effect of a study not
included in it is a new draw from the generative distribution \(
  \text{Student-}t(\nu,\ \beta,\ \tau_\beta).\)
Its predictive distribution, obtained by integrating the generative distribution
against the posterior of $(\beta, \tau_\beta)$, can be used in two ways. In the \emph{retrospective} case, we leave out one of the observed studies. Fixing the left-out study $\ell \in \{1,\dots,S\}$, we predict its effect from the rest, $\mathcal{D}_{-\ell}$,
\begin{equation}\label{eq:predictive-retro}
  p(\beta_\ell^{\mathrm{pred}} \mid \mathcal{D}_{-\ell})
  = \iint \text{Student-}t(3,\beta,\tau_\beta)\,
    p(\beta, \tau_\beta \mid \mathcal{D}_{-\ell})\, d\beta\, d\tau_\beta .
\end{equation}
In the \emph{prospective} case, we instead use all $S$ studies, $\mathcal{D}$, to predict the effect of a future study,
\begin{equation}\label{eq:predictive-pro}
  p(\beta_{\text{new}} \mid \mathcal{D})
  = \iint \text{Student-}t(3,\beta,\tau_\beta)\,
    p(\beta, \tau_\beta \mid \mathcal{D})\, d\beta\, d\tau_\beta .
\end{equation}

Because $\tau_\beta$ is random, in both cases the inner law is averaged over the full posterior of the heterogeneity, so the predictive distribution has no closed form. Fixing $\tau_\beta$ would remove the integration in Equations \eqref{eq:predictive-retro}--\eqref{eq:predictive-pro} and, with normal components, return the closed-form normal predictive of Pawel and Held. In our case the predictive distribution is approximated by simulation: for each posterior sample $t = 1,\dots,T$ we draw an independent deviation $\xi^{(t)} \sim \text{Student-}t(3, 0, 1)$ and set
\begin{equation}\label{eq:sim}
  \beta_\ell^{\mathrm{pred (t)}} = \beta^{(t)}_{-\ell} + \tau^{(t)}_{\beta,-\ell}\, \xi^{(t)}
  \qquad\text{or}\qquad
  \beta_{\text{new}}^{(t)} = \beta^{(t)} + \tau_\beta^{(t)}\, \xi^{(t)},
\end{equation}
using draws from $p(\beta, \tau_\beta \mid \mathcal{D}_{-\ell})$ in the retrospective case and from $p(\beta, \tau_\beta \mid \mathcal{D})$ in the prospective one.

In the retrospective case, the predicted effect $\beta_\ell^{\mathrm{pred}}$ is compared with the observed effect $\beta_\ell$, the latter obtained by fitting the model in its independence limit, to the data of study $\ell$ alone. Since the observed effect is fitted using only study $\ell$'s data and the predicted one uses the other studies, the two are computed from separate data, with a common prior; we treat their draws as independent and compare them by pairing the MCMC draws.

The retrospective question is the analog, in parameter space and using our metrics, of the out-of-sample question that \citet{pawel2020probabilistic} address in the empirical space. However, it should not be considered a replicability question in a strict sense: we are not assessing the reliability of the conclusions about the effect of interest. It can rather be seen as a way to gauge how far the studies depart from exchangeability, by computing the replicability probabilities between the observed effect and the one predicted from the generative distribution from which that study is assumed to derive. 

The prospective use serves yet another purpose. In the prospective case there is no observed study to compare against. The predictive distribution of $\beta_{\text{new}}$ describes what to expect from a study that has not yet been run, and the probability $P(|\beta_{\text{new}}| > \varepsilon)$ measures how likely such a study is to show an effect of practical relevance before any data are collected. This is useful for planning a new study. A failure to replicate is often diagnosed only afterwards, once a disappointing result is explained by low power or a small sample; a design analysis can and should be carried out prospectively \citep{gelman2014beyond}. The predictive lets the same framework anticipate that diagnosis in advance. A high predictive probability, for instance, indicates that the effect is likely to be of practical relevance in a new study as well, which is useful when one wants to confirm the conclusions about the effect of interest before applying a treatment in practice. When it is low, it is worth asking why before planning a new study: if it reflects the small number of studies observed so far, more evidence is worthwhile and would sharpen the prediction; if it reflects genuine heterogeneity across studies, no additional study will make the effect replicate. This distinction has practical implications: in clinical meta-analyses, for example, where non-negligible between-study heterogeneity is common \citep{inthout2016plea}, recognizing it would rule out planning a new study in the hope of finding the effect merely because more data would then be available, and would instead point to accounting for that heterogeneity in the analysis.

\section{Empirical Application}\label{sec:empirical}
Having developed the framework and the metrics it supports, we now apply it to an empirical example, assessing the replicability of the effect of a genetic variant on gene expression, across human ancestral groups.

Each ancestral group plays the role of a study in the sense of Section~\ref{sec:fixedandrandomfactors}: its donors are a sample from a distinct target population, and the target populations are themselves drawn from a common super-population.

\subsection{eQTL Analysis}
We draw the empirical example from the Genotype--Tissue Expression (GTEx) project \citep{gtex2020gtex}, launched to characterize genetic effects on the transcriptome across human tissues and to link these regulatory mechanisms to trait and disease associations. The object of study is an expression quantitative trait locus (eQTL): a genetic variant whose alleles are associated with the expression level of one or more genes, whether close by on the genome (a \emph{cis}-eQTL) or distant from it (a \emph{trans}-eQTL). 

Many variants influence complex traits not directly but through gene expression, and often not of the nearest gene: a single \emph{trans}-eQTL can affect genes and shape entire regulatory networks, typically by acting through transcription factors \citep{vosa2021large}.

A well-known example involves the transcription factor \textit{KLF14}. Variants near the \textit{KLF14} locus are robustly associated with type 2 diabetes and HDL cholesterol levels \citep{voight2010twelve, teslovich2010biological}. These variants form a \emph{cis}-regulatory haplotype, a group of highly correlated single nucleotide polymorphisms (SNPs) associated with \textit{KLF14} expression in adipose tissue, and are at the same time a \emph{trans}-eQTL of a broad downstream network, since \textit{KLF14} in turn acts as its master \emph{trans}-regulator \citep{small2011identification}. In this analysis we focus on the \emph{cis} effect: the effect of the lead variant rs4731702 of the haplotype on \textit{KLF14} expression \citep{small2018regulatory, kong2009parental}. This effect has so far been characterized almost exclusively in European-ancestry cohorts \citep{small2018regulatory}, leaving open whether it replicates across ancestral groups, a question linked to the broader literature about cross-ancestry eQTL studies \citep{stranger2012patterns, mogil2018genetic, shang2020genetic, kachuri2023gene}. In this example, the generative effect represents the effect of the genetic variant on \textit{KLF14} across all ancestral groups, while the study-specific effects describe how that influence is realized within each ancestry. A non-zero generative effect does not guarantee that the effect recurs within every ancestry, and the converse holds as well: agreement among a few ancestries need not reflect a non-zero effect shared across the wider super-population.

Before proceeding with the analysis, it is worth clarifying the effect of interest. The replicability we quantify concerns the effect as measured at rs4731702. Because rs4731702 is a marker correlated with the causal variant, and this correlation may differ across ancestral groups, we interpret the observed between-ancestry differences as differences in the measured effect, not as evidence of a different underlying regulatory mechanism; \citet{kachuri2023gene, gay2020impact, mogil2018genetic} show that most such differences are attributable to differences in allele frequency and linkage-disequilibrium patterns across populations. Separating genuine heterogeneity from differences in how well the marker tags the causal variant would require further genetic analysis beyond the scope of the present work. Moreover, \textit{KLF14} is expressed only from the maternally inherited allele \citep{small2011identification, small2018regulatory}. Since the genotypes in GTEx do not carry parent-of-origin information, the estimated effect is, in each ancestral group, half the true one, and should be interpreted accordingly.
  
\subsection{Data Acquisition and Preprocessing}
The analysis draws on two data releases of the GTEx project. 
Genotypes were taken from the GTEx v8 whole-genome sequencing release (838 individuals) and expression from the GTEx v10 subcutaneous-adipose RNA-seq release (714 donors). Restricting to donors present in both releases yields the analysis sample of 628. These datasets are available under controlled access and cannot be redistributed; the R code implementing the analysis, together with a synthetic data set, is available at \url{https://github.com/esteralongi/The-Quantification-Of-Replicability}.

Expression is quantified as raw read counts. Genes expressed at a negligible level are removed, retaining only those with more than 0.1 transcripts per million in at least a fifth of the samples and at least six reads in at least a fifth of the samples, following the GTEx criteria (\url{https://www.gtexportal.org/home/}). Expression values are normalized between samples using the median of ratios size factors of DESeq2 \citep{love2014moderated}, which rescale each sample to correct for differences in sequencing depth and library composition between donors. The normalized counts are strongly right-skewed and bounded below at zero, the floor corresponding to the absence of detected transcripts; expression is thus square-root transformed, the variance-stabilizing transformation for count data \citep{bartlett1936square}.

The genotype is coded additively as the number of alternative alleles carried by each donor: zero for the reference homozygote, one for the heterozygote and two for the alternative homozygote. 

In our model (Equation~\ref{eq:model}), $Y_{si}$ is the transformed expression of the gene \textit{KLF14} in donor $i$ of ancestry $s$, $X_{si}$ is the genotype of donor $i$ of ancestry $s$ at the regulatory variant rs4731702 in terms of number of alternative alleles, and the study-specific effect $\beta_s$ represents the change in transformed gene expression per additional alternative allele in the ancestral group $s$.

The empirical illustration deliberately adopts the simplest sufficient model, an intercept and the genotype, to keep attention on the replicability machinery rather than on eQTL modeling; technical covariates (e.g. expression factors, sequencing platform, donor sex) can be added directly and would mainly sharpen the residual scale $\sigma_s$, without altering the framework.

\subsection{Defining the Studies through Genetic Ancestry}
The ancestral groups are defined from the self-reported ancestry labels recorded for each donor in the GTEx v8 phenotype release. We partition the donors into $S = 3$ clusters that comprise 78, 531 and 19 donors and correspond to individuals reporting African American, European American and a third combined group, respectively. The third group, which we label ``Other'', pools the donors who reported Asian American race (12 donors) or Hispanic or Latino ethnicity (7 donors); we merge these two categories because their sample sizes are too small to support a study-specific estimate.

Although the three resulting studies are markedly unequal in size, we deliberately retain all three. With only 19 donors, a study-specific estimate is uncertain, but Bayesian analysis explicitly carries that uncertainty in the width of the posterior distribution.

\subsection{Empirical Bayes Prior Elicitation}\label{sec:prior_elicitation}
The priors of the generative model are elicited from data, through an empirical Bayes procedure. The human genome contains more than a thousand transcription factors regulated by nearby variants through broadly analogous mechanisms, so a large pool of comparable regulatory relationships is available to inform the priors for the relationship under study. The target pair rs4731702--KLF14 is excluded from this elicitation pool, so the hyperparameters are informed by analogous relationships rather than by the target's data.

We assemble the pool of variant--transcription-factor pairs to match, as far as possible, the regulatory mechanism of the target pair, as follows. Starting from the catalog of $1{,}639$ human transcription factors of \citet{lambert2018human}, we retain the significant GTEx eQTLs whose associated gene is a transcription factor, yielding 941 transcription factors. We then obtain signed transcription factor--target gene interactions from the CollecTRI collection \citep{muller2023expanding}, accessed through the \texttt{collectri} function of the \texttt{OmnipathR} package \citep{turei2021integrated}. We retain interactions flagged as directed, that is, documented regulatory relationships rather than mere protein--protein contacts, and supported either by at least two of the constituent resources or by at least two literature references, information reported in the \texttt{collectri} function. Because our target transcription factor, \textit{KLF14}, is a \emph{trans}-regulator, we keep only transcription factors that likewise regulate target genes on different chromosomes. The effect we model for every pair is the \emph{cis-eQTL} of the variant on the transcription factor's expression; that is, we regress the transcription factor's expression on the variant's genotype, exactly as for rs4731702 on \textit{KLF14}; the target gene enters only through this \emph{trans-regulator} selection and is not itself modeled. Finally, to ensure that the pool reflects the biological context of the target relationship, we rank candidates by two matched criteria: the distance in allele frequency from the target variant, and the distance in mean and variance of transcription factor expression from the target transcription factor. We retain one pair per variant and per transcription factor to obtain independent pairs, keeping the 280 closest as the elicitation pool.

For each pair $j$ in the pool we fit, separately within each ancestry $s$, a simple regression of the transcription factor's expression on genotype, obtaining an estimate $\hat{\beta}_{sj}$ of the regulatory effect with standard error $SE_{sj}$, together with the intercept $\hat{\alpha}_{sj}$ and residual scale $\hat{\sigma}_{sj}$. Of the 280 initial pairs, 227 yield a finite slope estimate with a standard error in all three ancestral groups and enter the elicitation; the remaining pairs fail in at least one ancestral group, typically because the variant is monomorphic in that ancestral group (all donors carry the same genotype, so its effect on expression cannot be estimated). The elicitation then proceeds in two steps.

\subsubsection{Step 1: Choice of the prior distributions.}
For each pair $j$, we summarize every quantity across the $S=3$ ancestral groups by an inverse-variance weighted mean and by the between-group heterogeneity. We estimate the heterogeneity with the DerSimonian--Laird estimator \citep{dersimonian1986meta}, which subtracts the expected sampling contribution and uses a bias-corrected denominator. For the effect, with precision weights $w^\beta_{sj}=1/SE(\hat \beta_{sj})^2$, we compute
\begin{align}
  \beta_j &= \frac{\sum_{s} w^\beta_{sj}\,\hat{\beta}_{sj}}{\sum_{s} w^\beta_{sj}},
\end{align}
\begin{align}
  \tau_{\beta,j}^2 &= \max\!\left(0,\ \frac{Q_{\beta,j}-(S-1)}
        {\sum_{s} w^\beta_{sj}-\sum_{s} (w^\beta_{sj})^2/\sum_{s} w^\beta_{sj}}\right),
\qquad
\nonumber\\
    \text{where} \qquad Q_{\beta,j} &= \sum_{s} w^\beta_{sj}\,(\hat{\beta}_{sj}-\beta_j)^2 .
    \nonumber
\end{align}
The intercepts follow the same construction with $\hat{\alpha}_{sj}$ and weights $w^\alpha_{sj}=1/SE(\hat \alpha_{sj})^2$,
\begin{align}
  \alpha_j &= \frac{\sum_{s} w^\alpha_{sj}\,\hat{\alpha}_{sj}}{\sum_{s} w^\alpha_{sj}},
\end{align}
\begin{align}
  \tau_{\alpha,j}^2 &= \max\!\left(0,\ \frac{Q_{\alpha,j}-(S-1)}
        {\sum_{s} w^\alpha_{sj}-\sum_{s} (w^\alpha_{sj})^2/\sum_{s} w^\alpha_{sj}}\right),
\qquad
\nonumber\\
    \text{where} \quad Q_{\alpha,j} &= \sum_{s} w^\alpha_{sj}\,(\hat{\alpha}_{sj}-\alpha_j)^2 .
        \nonumber
\end{align}
For the residual scales, a residual standard deviation estimate on $d_{sj}$ degrees of freedom has sampling variance $\widehat{\mathrm{Var}}(\hat{\sigma}_{sj}) =\hat{\sigma}_{sj}^2/(2 d_{sj})$ \citep[Part~Two, \S7, p.~135]{kenney1939stat2}; therefore, we use the corresponding precision weights $\tilde{w}_{sj}=2 d_{sj}/\hat{\sigma}_{sj}^2$,
\begin{align}
  \sigma_j &= \frac{\sum_{s} \tilde{w}_{sj}\,\hat{\sigma}_{sj}}{\sum_{s} \tilde{w}_{sj}},
\end{align}
\begin{align}
  \tau_{\sigma,j}^2 &= \max\!\left(0,\ \frac{Q_{\sigma,j}-(S-1)}
        {\sum_{s} \tilde{w}_{sj}-\sum_{s} \tilde{w}_{sj}^2/\sum_{s} \tilde{w}_{sj}}\right),
\qquad
\nonumber\\
    \text{where} \quad Q_{\sigma,j} &= \sum_{s} \tilde{w}_{sj}\,(\hat{\sigma}_{sj}-\sigma_j)^2 .
    \nonumber
\end{align}

Across the 227 pairs, the empirical distribution of each pooled quantity ($\beta_j$, $\alpha_j$, $\sigma_j$, and the heterogeneities $\tau_{\beta,j}$, $\tau_{\alpha,j}$, $\tau_{\sigma,j}$) has markedly heavier tails than a Gaussian, so we adopt Student-$t$ priors with $\nu = 3$ degrees of freedom. The effects $\beta_j$ spread on both sides of zero, so the prior on $\beta$ is an untruncated Student-$t(3)$, whereas the intercepts $\alpha_j$, the residual scales $\sigma_j$ and the three heterogeneities $\tau_{\beta,j},\tau_{\alpha,j},\tau_{\sigma,j}$ are non-negative, so the priors on $\alpha$, $\sigma$ and $\tau_\beta,\tau_\alpha, \tau_\sigma$ are Student-$t(3)$ truncated to $[0,\infty)$. For the heterogeneity parameters the truncated Student-$t$ is preferable not only for its empirical fit but also because, as we noted in Section~\ref{sec:generativemodel}, the prior on $\tau$ encodes how strongly we believe the analysis is a replication rather than an extension through the mass it places near zero, and unlike the Gamma and log-normal densities, the truncated Student-$t$ retains a finite positive density at zero, admitting the replication limit $\tau \to 0$.

\subsubsection{Step 2: Choice of the hyperparameters.}
For each quantity, we summarize its pair-level distribution across the 227 pairs by its empirical mean and standard deviation, and set the parameters of the corresponding Student-$t(3)$ prior so that the prior reproduces these two moments. For the effect $\beta$, the location equals the empirical mean and the scale follows from the closed-form relation between the standard deviation and the scale of an untruncated Student-t distribution. Let $t_{\mu,\eta,\nu} = \text{Student-t}(\nu)$ with location $\mu$ and scale $\eta$, $\mathrm{SD}=\eta\sqrt{\nu/(\nu-2)}$:
\begin{align}
  \mu_{\beta} &= \mathrm{Mean}(\beta_j),
  & \eta_{\beta} &= \sqrt{\tfrac{\nu-2}{\nu}}\,\mathrm{SD}(\beta_j)
      = \sqrt{\tfrac{1}{3}}\,\mathrm{SD}(\beta_j).
\end{align}
For the five parameters restricted to the positive half-line ($\alpha$, $\sigma$, $\tau_\beta$, $\tau_\alpha$, $\tau_\sigma$) this closed form no longer applies: truncation at zero changes both the mean and the variance of the distribution, so the untruncated relation overstates the scale and the location no longer coincides with the mean. For each such quantity, among the Student-$t(\nu)$ distributions truncated to $[0,\infty)$, we select the location $\mu$ and scale $\eta$ whose first two moments best match the empirical ones. Let $t^{+}_{\mu,\eta,\nu}$ denote a Student-$t(\nu)$ with location $\mu$ and scale $\eta$ truncated to $[0,\infty)$. The chosen hyperparameters make the truncated moments as close as possible to the empirical ones,
\begin{align}
  \mathrm{Mean}\!\big(t^{+}_{\mu_{\alpha},\eta_{\alpha},\nu}\big) &\approx \mathrm{Mean}(\alpha_j),
    & \mathrm{SD}\!\big(t^{+}_{\mu_{\alpha},\eta_{\alpha},\nu}\big) &\approx \mathrm{SD}(\alpha_j),\\
  \mathrm{Mean}\!\big(t^{+}_{\mu_{\sigma},\eta_{\sigma},\nu}\big) &\approx \mathrm{Mean}(\sigma_j),
    & \mathrm{SD}\!\big(t^{+}_{\mu_{\sigma},\eta_{\sigma},\nu}\big) &\approx \mathrm{SD}(\sigma_j),\\
  \mathrm{Mean}\!\big(t^{+}_{\mu_{\tau_\beta},\eta_{\tau_\beta},\nu}\big) &\approx \mathrm{Mean}(\tau_{\beta,j}),
    & \mathrm{SD}\!\big(t^{+}_{\mu_{\tau_\beta},\eta_{\tau_\beta},\nu}\big) &\approx \mathrm{SD}(\tau_{\beta,j}),\\
  \mathrm{Mean}\!\big(t^{+}_{\mu_{\tau_\alpha},\eta_{\tau_\alpha},\nu}\big) &\approx \mathrm{Mean}(\tau_{\alpha,j}),
    & \mathrm{SD}\!\big(t^{+}_{\mu_{\tau_\alpha},\eta_{\tau_\alpha},\nu}\big) &\approx \mathrm{SD}(\tau_{\alpha,j}),\\
  \mathrm{Mean}\!\big(t^{+}_{\mu_{\tau_\sigma},\eta_{\tau_\sigma},\nu}\big) &\approx \mathrm{Mean}(\tau_{\sigma,j}),
    & \mathrm{SD}\!\big(t^{+}_{\mu_{\tau_\sigma},\eta_{\tau_\sigma},\nu}\big) &\approx \mathrm{SD}(\tau_{\sigma,j}).
\end{align}
Together, these locations and scales fully specify the generative model: the effect is assigned the untruncated prior $\beta \sim t_{\mu_\beta,\eta_\beta,3}$, while the intercept, the residual scale, and the three heterogeneities are assigned the truncated priors $t^{+}_{\mu_\alpha,\eta_\alpha, 3}$, $t^{+}_{\mu_\sigma,\eta_\sigma, 3}$, $t^{+}_{\mu_{\tau_\beta},\eta_{\tau_\beta}, 3}$, $t^{+}_{\mu_{\tau_\alpha},\eta_{\tau_\alpha}, 3}$, and $t^{+}_{\mu_{\tau_\sigma},\eta_{\tau_\sigma}, 3}$, with the numerical values reported in Table~\ref{tab:priors}.

\begin{table}[!ht]
\centering
\caption{Empirical Bayes hyperparameters elicited from the pool of 227 regulatory pairs. $\mu,\eta$ are the location and scale of the Student-$t(3)$ prior on the parameter ($\beta$ untruncated; $\alpha,\sigma$ truncated to $[0,\infty)$), and $\mu_\tau,\eta_\tau$ are the location and scale of the truncated Student-$t(3)$ prior on the corresponding heterogeneity parameters $\tau_\beta,\tau_\alpha,\tau_\sigma$.}
\label{tab:priors}
\small
\begin{tabular}{lcccc}
\toprule
 & $\mu$ & $\eta$ & $\mu_\tau$ & $\eta_\tau$ \\
\midrule
Effect $\beta$        & $0.044$ & $0.63$ & $-1.05$ & $0.084$ \\
Intercept $\alpha$    & $34.90$  & $23.63$ & $-1.36$ & $0.11$  \\
Residual $\sigma$     & $4.88$  & $4.66$ & $-1.0$ & $0.080$ \\
\bottomrule
\end{tabular}
\end{table}

\subsection{A Threshold of Practical Relevance}\label{epsilon_choice}
We propose to assess the replicability of effects with respect to a threshold $\varepsilon$, separating effects of practical relevance from those practically equivalent to the null. When sufficient a priori information is available to set $\varepsilon$ in a way that is independent of the data, the value should be set from that scientific judgment. Yet, for the regulatory effect of rs4731702 on \textit{KLF14}, no external anchor is available that is both commensurable with our measurement scale and independent of our data. The association has been characterized in the same tissue in independent data \citep{small2018regulatory}, but on a different expression scale. More precisely, they corrected exon read counts for technical covariates using a mixed-effects model, and used the corrected residuals for all eQTL analyses. Thus, a threshold defined on that scale does not transfer to our effects, computed on gene expression normalized with DESeq2 and square-root transformed. Therefore, since a threshold of practical relevance cannot be specified a priori, we tie $\varepsilon$ to the estimated baseline on the model's scale. In particular, we define an effect as practically relevant when it is at least a fraction $p$ of the baseline expression $\alpha$, the normalized mean expression of \textit{KLF14} on the square-root scale in the genotype 0-class, the class corresponding to no copies of the variant. The threshold is therefore \(\varepsilon = p\,\alpha .\)
We take $p = 0.10$; with $\alpha = 8.61$ this gives $\varepsilon = 0.86$: an effect at the threshold changes expression on the raw count scale by $(1+p)^2 - 1 = (1+0.10)^2 - 1 = 21\%$ per additional allele. Since $\varepsilon$ is on the scale of the estimated effects, it corresponds to half of the biologically relevant effect.

A $p$-value conflates the magnitude of the effect with the sample size, so on its own it cannot distinguish a genuinely relevant effect from a negligible one made significant merely by a large sample. By contrast, the threshold so constructed is set according to the scientific relevance of the effect; it is therefore expressed on the scale of the coefficients and is independent of the sample size.

\subsection{Results}\label{sec:results}
We fit the hierarchical model specified in Equation~\ref{eq:model} to the three ancestral groups European, African and ``Other'', with the empirically elicited priors and the threshold $\varepsilon = 0.86$. The estimated effect across the comparable regulatory pairs is small on average ($\mu_\beta = 0.044$, see Table~\ref{tab:priors}) but highly variable: the central 95\% interval of $t_{\mu_\beta,\eta_\beta, 3}$, $[-1.96, 2.05]$ extends well beyond $\varepsilon = 0.86$ in both directions, so the prior neither assumes an effect of practical equivalence to the null nor rules out one of practical relevance. A single regression of \textit{KLF14} expression on rs4731702, using all donors, yields an estimated effect of $0.94$, which falls within this interval. 

Posterior sampling is carried out in Stan \citep{carpenter2017stan} via the \texttt{rstan} interface, running four Hamiltonian Monte Carlo chains of 3000 iterations each: the first 1500 discarded as warmup, for 6000 posterior draws in total. Sampling is stable, with split-$\widehat{R} = 1.0$ for all parameters and no divergent transitions.

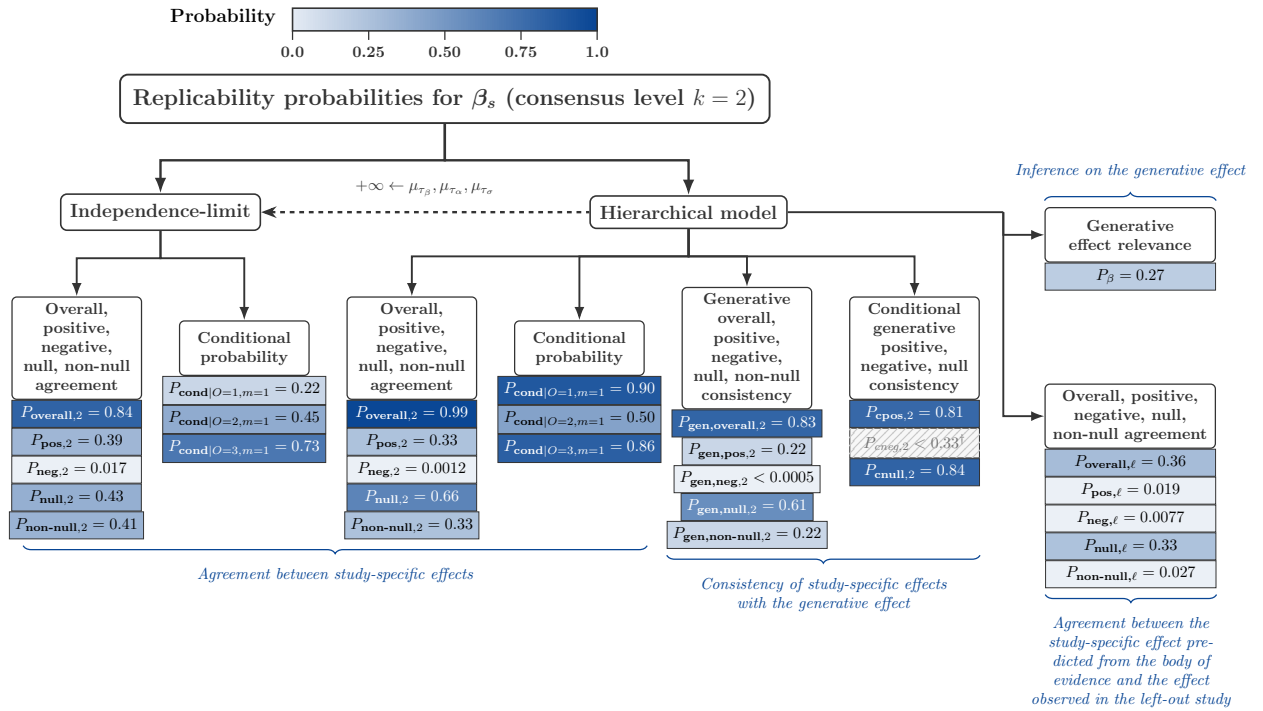
\begin{figure}[!ht] 
\centering
\definecolor{DeepBlue}{RGB}{8, 69, 148} 
\definecolor{graytext}{gray}{0.2}
\resizebox{\textwidth}{!}{%
   \begin{tikzpicture}[>=Stealth, node distance=0pt, xscale=1.12]
    
    \tikzstyle{root} = [draw=graytext, fill=white, rounded corners=5pt, inner sep=8pt, font=\Large\bfseries\color{graytext}, align=center, line width=1.2pt]
    \tikzstyle{branch} = [draw=graytext, fill=white, rounded corners=4pt, inner sep=6pt, font=\large\bfseries\color{graytext}, align=center, line width=1pt]
    \tikzstyle{box} = [draw=graytext, fill=white, rounded corners=3pt, text width=2.6cm, minimum height=1.2cm, font=\small\bfseries\color{graytext}, align=center, line width=0.8pt]
    
    \tikzstyle{heat} = [draw=graytext, minimum width=2.86cm, minimum height=0.55cm, font=\small\bfseries, line width=0.5pt, align=center]
    \tikzstyle{heatW} = [heat, text=white] 
    \tikzstyle{heatB} = [heat, text=black] 
    \usetikzlibrary{patterns}
    \tikzstyle{heatU} = [heat, fill=black!3, text=black!55, font=\small\itshape, postaction={pattern=north east lines, pattern color=black!22}]
    
    \tikzstyle{qtext} = [font=\small\itshape, text=DeepBlue, align=center]
    \node[left, font=\large\color{black}] at (-3.2, 3.8) {\textbf{Probability}};
    \shade[left color=DeepBlue!10, right color=DeepBlue, draw=graytext, line width=1pt] (-3, 3.5) rectangle (3, 4);
    \draw[thick, graytext] (-3, 3.5) -- (-3, 3.3) node[below, font=\small\bfseries\color{graytext}] {0.0};
    \draw[thick, graytext] (-1.5, 3.5) -- (-1.5, 3.3) node[below, font=\small\bfseries\color{graytext}] {0.25};
    \draw[thick, graytext] (0, 3.5) -- (0, 3.3) node[below, font=\small\bfseries\color{graytext}] {0.50};
    \draw[thick, graytext] (1.5, 3.5) -- (1.5, 3.3) node[below, font=\small\bfseries\color{graytext}] {0.75};
    \draw[thick, graytext] (3, 3.5) -- (3, 3.3) node[below, font=\small\bfseries\color{graytext}] {1.0};
    \node[root] (root) at (0, 2) {Replicability probabilities for $\boldsymbol{\beta_s}$ (consensus level $k=2$)};
    
    \node[branch] (indep) at (-5.6, -0.5) {Independence-limit};
    \node[branch] (dep)   at (4.8, -0.5) {Hierarchical model};
    \node[box] (i_agree)  at (-7.25, -3.5) {Overall, positive, negative, null, non-null agreement};
    \node[box] (i_cond)   at (-3.95, -3.5) {Conditional probability};
    
    \node[box] (d_agree1) at (-0.65, -3.5) {Overall, positive, negative, null, non-null agreement};
    \node[box] (d_cond1)  at (2.65, -3.5)  {Conditional probability};
    \node[box] (d_agree2) at (5.95, -3.5)  {Generative overall, positive, negative, null, non-null consistency};
    \node[box] (d_cond2)  at (9.25, -3.5)  {Conditional generative positive, negative, null consistency};
    \node[box, text width=3.5cm] (d_meta) at (13.5, -1) {Generative effect relevance};
    \node[box, text width=3.5cm] (d_body) at (13.5, -5) {Overall, positive, negative, null, non-null agreement};
    \node[heatW, fill=DeepBlue!84, below=0pt of i_agree] (ia1) {$P_{\text{overall},2}=0.84$};
    \node[heatB, fill=DeepBlue!39, below=0pt of ia1]     (ia2) {$P_{\text{pos},2}=0.39$};
    \node[heatB, fill=DeepBlue!8, below=0pt of ia2]     (ia3) {$P_{\text{neg},2}=0.017$};
    \node[heatB, fill=DeepBlue!43, below=0pt of ia3]     (ia4) {$P_{\text{null},2}=0.43$};
    \node[heatB, fill=DeepBlue!41, below=0pt of ia4]     (ia5) {$P_{\text{non-null},2}=0.41$};
    \node[heatB, fill=DeepBlue!22, below=0pt of i_cond] (ic1) {$P_{\text{cond}\mid O = 1, m = 1}=0.22$};
    \node[heatB, fill=DeepBlue!45, below=0pt of ic1]    (ic2) {$P_{\text{cond}\mid O = 2, m = 1}=0.45$};
    \node[heatW, fill=DeepBlue!73, below=0pt of ic2]    (ic3) {$P_{\text{cond}\mid O = 3, m = 1}=0.73$};
    \node[heatW, fill=DeepBlue!99, below=0pt of d_agree1] (da1) {$P_{\text{overall},2}=0.99$};
    \node[heatB, fill=DeepBlue!33, below=0pt of da1]      (da2) {$P_{\text{pos},2}=0.33$};
    \node[heatB, fill=DeepBlue!8, below=0pt of da2]      (da3) {$P_{\text{neg},2}=0.0012$};
    \node[heatW, fill=DeepBlue!66, below=0pt of da3]      (da4) {$P_{\text{null},2}=0.66$};
    \node[heatB, fill=DeepBlue!33, below=0pt of da4]      (da5) {$P_{\text{non-null},2}=0.33$};
    \node[heatW, fill=DeepBlue!90, below=0pt of d_cond1] (dc1) {$P_{\text{cond}\mid O = 1, m = 1}=0.90$};
    \node[heatB, fill=DeepBlue!50, below=0pt of dc1]     (dc2) {$P_{\text{cond}\mid O = 2, m = 1}=0.50$};
    \node[heatW, fill=DeepBlue!86, below=0pt of dc2]     (dc3) {$P_{\text{cond}\mid O = 3, m = 1}=0.86$};
    \node[heatW, fill=DeepBlue!83, below=0pt of d_agree2] (d2a1) {$P_{\text{gen,overall},2}=0.83$};
    \node[heatB, fill=DeepBlue!22, below=0pt of d2a1]     (d2a2) {$P_{\text{gen,pos},2}=0.22$};
    \node[heatB, fill=DeepBlue!8, below=0pt of d2a2]     (d2a3) {$P_{\text{gen,neg},2}<0.0005$};
    \node[heatW, fill=DeepBlue!61, below=0pt of d2a3]     (d2a4) {$P_{\text{gen,null},2}=0.61$};
    \node[heatB, fill=DeepBlue!22, below=0pt of d2a4]     (d2a5) {$P_{\text{gen,non-null},2}=0.22$};
    \node[heatW, fill=DeepBlue!81, below=0pt of d_cond2] (d2c1) {$P_{\text{cpos},2}=0.81$};
    \node[heatU, below=0pt of d2c1]    (d2c2) {$P_{\text{cneg},2}<0.33^{\dagger}$};
    \node[heatW, fill=DeepBlue!84, below=0pt of d2c2]    (d2c3) {$P_{\text{cnull},2}=0.84$};
    \tikzstyle{heat_wide} = [heat, minimum width=3.76cm] 
    \node[heat_wide, text=black, fill=DeepBlue!27, below=0pt of d_meta] (lm) {$P_{\beta}=0.27$};
    \node[heat_wide, text=black, fill=DeepBlue!36, below=0pt of d_body] (lf1) {$P_{\text{overall,}\ell}=0.36$};
    \node[heat_wide, text=black, fill=DeepBlue!8,  below=0pt of lf1]    (lf2) {$P_{\text{pos,}\ell}=0.019$};
    \node[heat_wide, text=black, fill=DeepBlue!8, below=0pt of lf2]    (lf3) {$P_{\text{neg,}\ell}=0.0077$};
    \node[heat_wide, text=black, fill=DeepBlue!33, below=0pt of lf3]    (lf4) {$P_{\text{null,}\ell}=0.33$};
    \node[heat_wide, text=black, fill=DeepBlue!8,  below=0pt of lf4]    (lf5) {$P_{\text{non-null,}\ell}=0.027$};
    \draw[decorate, decoration={brace, amplitude=5pt, mirror}, thick, DeepBlue] 
        ([xshift=8pt, yshift=-4pt]ia5.south west) -- ([xshift=-8pt, yshift=-4pt]dc3.south east |- ia5.south) 
        node[midway, below=10pt, qtext] {Agreement between study-specific effects};
    \draw[decorate, decoration={brace, amplitude=5pt, mirror}, thick, DeepBlue] 
        ([yshift=-4pt]d2a5.south west) -- ([yshift=-4pt]d2c3.south east |- d2a5.south) 
        node[midway, below=10pt, qtext] {Consistency of study-specific effects \\ with the generative effect};
    \draw[decorate, decoration={brace, amplitude=5pt}, thick, DeepBlue] 
        ([yshift=4pt]d_meta.north west) -- ([yshift=4pt]d_meta.north east) 
        node[midway, above=10pt, qtext] {Inference on the generative effect};
    \draw[decorate, decoration={brace, amplitude=5pt, mirror}, thick, DeepBlue] 
        ([yshift=-4pt]lf5.south west) -- ([yshift=-4pt]lf5.south east)
        node[midway, below=10pt, qtext, text width=4.5cm] {Agreement between the study-specific effect predicted from the body of evidence and the effect observed in the left-out study};
    
    \draw[-Latex, line width=1.5pt, graytext] (root.south) -- +(0,-0.8) -| (indep.north);
    \draw[-Latex, line width=1.5pt, graytext] (root.south) -- +(0,-0.8) -| (dep.north);
    \draw[-Latex, line width=1.2pt, graytext] (indep.south) -- +(0,-0.6) -| (i_agree.north);
    \draw[-Latex, line width=1.2pt, graytext] (indep.south) -- +(0,-0.6) -| (i_cond.north);
    \draw[-Latex, line width=1.2pt, graytext] (dep.south) -- +(0,-0.6) -| (d_agree1.north);
    \draw[-Latex, line width=1.2pt, graytext] (dep.south) -- +(0,-0.6) -| (d_cond1.north);
    \draw[-Latex, line width=1.2pt, graytext] (dep.south) -- +(0,-0.6) -| (d_agree2.north);
    \draw[-Latex, line width=1.2pt, graytext] (dep.south) -- +(0,-0.6) -| (d_cond2.north);
    \draw[line width=1.2pt, graytext] (dep.east) -- (11.0, -0.5); 
    \draw[-Latex, line width=1.2pt, graytext] (11.0, -0.5) |- (d_meta.west); 
    \draw[-Latex, line width=1.2pt, graytext] (11.0, -0.5) |- (d_body.west); 
    \draw[<-, line width=1.5pt, graytext, dashed] (indep.east) -- node[above, font=\small\itshape\color{black}, yshift=0.2cm] {$+\infty \leftarrow \mu_{\tau_{\beta}}, \mu_{\tau_{\alpha}}, \mu_{\tau_{\sigma}}$} (dep.west);
    \end{tikzpicture}
}
\caption{Taxonomy of the replication posterior probabilities of the analyzed regulatory effect of the SNP rs4731702 on \textit{KLF14} in three ancestral groups, at the consensus level $k=2$ (conditional probabilities at $m=1$; predictive metrics evaluated retrospectively against the left-out African ancestral group). Heatmap intensity scales from white $(0.0)$ to deep blue $(1.0)$. The independence column corresponds to the limit $\mu_{\tau_\beta},\mu_{\tau_\alpha},\mu_{\tau_\sigma}\to\infty$. Probabilities of events not observed in any of the $6000$ draws are reported as the rule-of-three upper bound $3/T$, marked $<$; conditional estimates resting on fewer than $30$ conditioning draws are hatched and marked with a dagger ($\dagger$).}
\label{fig:taxonomy_tree}
\end{figure}

Figure~\ref{fig:taxonomy_tree} illustrates the complete set of metrics. The two columns contrast the independent analysis, the limit in which the heterogeneity scales $\tau_\beta,\tau_\alpha,\tau_\sigma$ are fixed at the 99.9th percentile of their respective prior distributions, with the hierarchical analysis. The generative effect relevance metric ($P_\beta$) is on the top-right, and the retrospective predictive block on the bottom-right. The taxonomy thus displays, in a single figure, what the data say about different replicability questions for the analyzed regulatory effect. Probabilities resting on fewer than 30 effective draws are flagged.

For every replication probability, we report the posterior estimate $\hat p$ together with a Monte Carlo standard error, $\mathrm{MCSE}(\hat p)=\sqrt{\hat p(1-\hat p)/\mathrm{ESS}}$, where $\mathrm{ESS}$ is the effective sample size (the conditional probabilities use the number of draws satisfying the conditioning event). $\mathrm{MCSE}(\hat p)$ is small throughout (the largest standard error across all metrics and models is $0.011$). Probabilities estimated as zero correspond to events not observed in any posterior draw, the negative-agreement metrics. This is not a chance result. The European ancestral group has a posterior mean effect of $0.93$ ($95\%$ CI $(0.53, 1.31)$), above $\varepsilon$, whereas the African and ``Other'' groups have posterior mean effects of practical equivalence to the null ($0.50$, $95\%$ CI $(-0.57, 1.25)$; and $0.63$, $95\%$ CI $(-0.44, 1.49)$, respectively). All three posterior means are positive, and almost no posterior draws place an effect below $-\varepsilon$. Consequently, there are very few draws in which two study-specific effects are simultaneously below $-\varepsilon$, and none in which all three are. The conditional generative negative consistency probability $P_{\text{cneg},2}$ likewise rests on few draws (9), since the posterior of the generative effect lies almost entirely above $-\varepsilon$ (posterior mean $0.61$, $95\%$ CI $(-0.27, 1.25)$).
All draw counts, ESSs and $\mathrm{MCSE}(\hat p)$ values are given in the Supplementary Material.

\subsubsection{Inference on the generative effect.}
The generative effect is of practical relevance with posterior probability $P_\beta = P(|\beta|>0.86) = 0.27$. Taken on its own, this figure tells us only that the generative effect is of practical relevance with probability little more than one in four and the width of the posterior distribution ($95\%$~CI~$(-0.27, 1.25)$) conveys the surrounding uncertainty. It says nothing about the underlying studies from which it arises. To read the deeper meaning this figure carries, we apply the proposed set of metrics.

\subsubsection{Agreement between study-specific effects.} 
We start by assessing the horizontal replicability by looking at all the metrics that involve just the comparison between study-specific effects of Section~\ref{sec:nogenerativemetrics}. The probability that all the study-specific effects are jointly above $\varepsilon$, below $-\varepsilon$, or of practical equivalence to the null, is $P_{\text{overall},3} = 0.43$. Under the milder consensus level $k=2$, the overall probability becomes $P_{\text{overall},2} = 0.99$, close to its prior of about $0.96$ (each effect falls in one of three regions, so two of three effects share a region unless the three fall in three different regions); thus, the data move $P_{\text{overall},2}$ only slightly. Its components are $P_{\text{pos},2} = 0.33$, $P_{\text{null},2} = 0.66$ and $P_{\text{neg},2}=0.0012$. The value each of these probabilities takes under the prior alone is reported in the Supplementary Material.

For each conditional probability, the conditioning ancestral group plays the role of the reference study $O$: $1$, $2$, $3$ correspond to the African, European and ``Other'' ancestral group, respectively. For these metrics, the value $m$ represents the minimum number of other studies required to corroborate the discovery of the reference study; the values reported here use $m=1$. Taking the European ancestral group as the reference study $O$, and restricting to the draws in which its effect is of practical relevance, the probability that at least one other ancestral group shows an effect of practical relevance and with the same sign is $P_{\text{cond}\mid O = 2, m = 1} = 0.50$: when the largest ancestral group shows an effect of practical relevance, the probability that at least one smaller group corroborates it is only about one half. Conditioning on the smaller ancestral groups, the conditional probabilities give higher values, $P_{\text{cond}\mid O = 1, m = 1} = 0.90$ and $P_{\text{cond}\mid O = 3, m = 1} = 0.86$: in the few draws in which the smaller ancestral groups show study-specific effects above the threshold, the European-specific effect is almost always above it as well.

To see how far these results depend on the elicited prior, we carried out a sensitivity analysis, varying the prior on $\tau_\beta$, the hyperparameter that controls the heterogeneity between the study-specific effects, and examining whether the replicability metrics remain stable. In particular, we set the prior median of $\tau_\beta$ to the 10th, 25th, 50th, 75th and 90th percentiles of the elicited prior. Table~\ref{tab:sens-tau} reports the replicability metrics obtained under these prior values.

\begin{table}[!ht]
\centering
\small
\caption{Sensitivity of the replicability probabilities ($k=2$) to the prior on $\tau_\beta$. The prior median $\mathrm{med}(\tau_\beta)$ is set to the 10th, 25th, 50th, 75th and 90th percentiles of the elicited empirical-Bayes prior; $\mathrm{med}(\tau_\beta)=0.28$ (marked $^{\ast}$) is the elicited value used in the main analysis. Probabilities of events not observed in any of the $6000$ posterior draws are reported as the rule-of-three upper bound $3/T$ ($<$); estimates whose conditioning event has fewer than $30$ draws are marked with a dagger ($\dagger$). All fits met a convergence criteria (no divergent transitions, $\hat R<1.02$ and minimum effective sample size $>100$).}
\label{tab:sens-tau}
\begin{tabular}{l r r r r r}
\toprule
$\mathrm{med}(\tau_\beta)$ & $0.038$ & $0.11$ & $0.28^{\ast}$ & $0.63$ & $1.2$ \\
\midrule
$P_{\mathrm{overall},2}$        & 1.0 & 1.0 & 0.99 & 0.98 & 0.98 \\
$P_{\mathrm{non\text{-}null},2}$ & 0.46 & 0.38 & 0.33 & 0.29 & 0.29 \\
$P_{\mathrm{pos},2}$            & 0.46 & 0.38 & 0.33 & 0.29 & 0.29 \\
$P_{\mathrm{null},2}$           & 0.55 & 0.62 & 0.66 & 0.70 & 0.69 \\
$P_{\mathrm{neg},2}$            & $<0.0005$ & $<0.0005$ & $0.0012$ & $<0.0005$ & $0.002$ \\
$P_{\mathrm{cond}\mid O=1, m=1}$ & 0.96 & 0.94 & 0.90 & 0.85 & 0.81 \\
$P_{\mathrm{cond}\mid O=2, m=1}$ & 0.79 & 0.61 & 0.50 & 0.41 & 0.41 \\
$P_{\mathrm{cond}\mid O=3, m=1}$ & 0.94 & 0.90 & 0.86 & 0.80 & 0.79 \\
$P_{\beta}$                     & 0.44 & 0.35 & 0.27 & 0.23 & 0.22 \\
$P_{\mathrm{gen,overall},2}$    & 0.92 & 0.88 & 0.83 & 0.79 & 0.77 \\
$P_{\mathrm{gen,non\text{-}null},2}$ & 0.41 & 0.31 & 0.22 & 0.16 & 0.15 \\
$P_{\mathrm{gen,pos},2}$        & 0.41 & 0.31 & 0.22 & 0.16 & 0.14 \\
$P_{\mathrm{gen,null},2}$       & 0.51 & 0.57 & 0.61 & 0.63 & 0.62 \\
$P_{\mathrm{gen,neg},2}$        & $<0.0005$ & $<0.0005$ & $<0.0005$ & $<0.0005$ & $<0.0005$ \\
$P_{\mathrm{cpos},2}$          & 0.93 & 0.88 & 0.81 & 0.71 & 0.69 \\
$P_{\mathrm{cnull},2}$         & 0.92 & 0.88 & 0.84 & 0.82 & 0.79 \\
$P_{\mathrm{cneg},2}$          & ---   & $0.43^{\dagger}$ & $<0.33^{\dagger}$ & $0.16^{\dagger}$ & $0.034^{\dagger}$ \\
\bottomrule
\end{tabular}
\end{table}

The overall agreement $P_{\mathrm{overall},2}$ varies between $1.0$ and $0.98$, and the study-level conditional probabilities for the two smaller groups are stable ($P_{\mathrm{cond}\mid O=1,m=1}$ from $0.96$ to $0.81$; $P_{\mathrm{cond}\mid O=3,m=1}$ from $0.94$ to $0.79$). Two quantities change more. First, $P_\beta$, the posterior probability that the generative effect is of practical relevance, decreases from $0.44$ to $0.22$: as $\tau_\beta$ grows, the posterior of $\beta$ spreads and places less mass in the region of practical relevance. Second, $P_{\mathrm{cond}\mid O=2,m=1}$, the probability that at least one other group corroborates the largest (European) group when it shows an effect of practical relevance, falls from $0.79$ to $0.41$: with more heterogeneity the model applies less shrinkage, so the two smaller groups are pulled less toward the common effect and corroborate the European one less often.

Since all the metrics illustrated above are computed under the hierarchical model and therefore partly reflect shrinkage rather than replicability alone, we now report them alongside the independence-limit metrics, obtained by fixing each heterogeneity scale $\tau_\beta,\tau_\alpha,\tau_\sigma$ to the 99.9th percentile of its prior distribution. In this case the posterior means of the three effects are as follows: the European group has a posterior mean effect of $1.08$ ($95\%$ CI $(0.67, 1.49)$), the African group $-0.29$ ($95\%$ CI $(-1.56, 0.97)$) and the Other group $0.64$ ($95\%$ CI $(-1.80, 3.08)$).

Without shrinkage, the African effect is no longer pulled toward the common generative effect and moves toward negative values, so it falls in the region of practical equivalence to the null less often: $P_{\mathrm{null},2}$, which in the hierarchical case was sustained mainly by the two smaller groups being shrunk toward the null, drops to $0.43$ (from $0.66$ under the hierarchical model), and $P_{\mathrm{neg},2}$ now rests on many more draws (100 rather than 7; see Table~\ref{tab:mcse-indep}). Accordingly, the conditional probability for the African group falls from $P_{\mathrm{cond}\mid O=1, m=1}=0.90$ under the hierarchical model to $0.22$: in most draws where the African effect is of practical relevance it lies below $-\varepsilon$, while the other two groups sit on the more positive side, so few draws remain in which they corroborate it.

\subsubsection{Consistency of the study-specific effects with the generative effect.}
The vertical replicability is assessed by looking at the consistency of the study-specific effects with the generative effect. The generative null consistency $P_{\mathrm{gen,null},2} = 0.61$ is large, and the conditional generative null consistency $P_{\mathrm{cnull},2} = 0.84$ indicates that among the draws in which the generative effect is of practical equivalence to the null, at least two ancestral groups agree, so the near-zero generative effect $\beta$ reflects a true consensus of small effects rather than a result of large opposing ones. The marginal probability $P(|\beta| \le 0.86) = 0.73$ is large enough for this conditional to rest on many draws, so the reading can be trusted.
 
\subsubsection{Agreement between the study-specific effect predicted from the body of evidence and the effect observed in the left-out study.}\label{subsec:prediction}
Finally, the same fitted model supports the predictive question: whether an ancestral group not among those analyzed would be consistent with a body of evidence. We treat the European ancestral group and the third one, composed of Asian and Hispanic ancestral groups, as the body of evidence $\mathcal{D}_{-\ell}$ and use this to predict the effect of the African ancestral group $\beta_\ell^{\mathrm{pred}}$.

The predicted effect for the left-out African ancestral group $\beta_\ell^{\mathrm{pred}}$ is compared with the observed African effect $\beta_\ell$ through the same agreement metrics used for pairs of observed studies; for example, we can compute the probability that the two effects are both above $\varepsilon$, both below $-\varepsilon$ or both of practical equivalence to the null ($P_{\text{overall,}\ell} = 0.36$) or the probability that they both have effects of practical relevance and in the same direction ($P_{\text{non-null,}\ell} = 0.027$). These values should be read with caution: even for a left-out study matching the prediction exactly, $P_{\text{overall,}\ell}$ cannot exceed about $0.52$, since the predictive distribution built from only two studies is wide; thus, the observed $0.36$ is close to this ceiling rather than low on a $0$--$1$ scale.

We can also predict the effect of a new study by treating all three studies as the body of evidence $\mathcal{D}$, as discussed in Section~\ref{sec:computational}. The probability that the effect of the new study is of practical relevance is $P(|\beta_{\text{new}}| > \varepsilon) = 0.38$.
 
\section{Calibrating the Metrics through Simulation}\label{sec:simulation}

\subsection{Motivation and Goals}
The empirical analysis returns the entire configuration of probabilities, but a probability read in isolation is hard to calibrate. We therefore present a simulation that varies the model's heterogeneity scales one at a time, anchored to our genomic application and using the same generative model, to see how the replicability metrics for $\beta$ respond. Keeping the generative model fixed and perturbing a single heterogeneity scale at a time lets us attribute the resulting changes in the metrics to that scale alone and read them against the empirical configuration.

Varying the heterogeneity of the effect itself, $\tau_\beta$, traces out the range the metrics can take as the studies move from low to high heterogeneity, giving a
scale against which the empirical probabilities can be read. For instance, they allow us to assess whether the metrics reflect near-perfect replicability when the between-study heterogeneity is nearly null, and they quantify the loss of agreement as the heterogeneity grows.
Varying the heterogeneity of the nuisance components, the intercepts
($\tau_\alpha$) and the residual scales ($\tau_\sigma$), shows whether, and how much, the replicability analysis for $\beta$ is disturbed when these components differ across studies.

\subsection{Empirically Grounded Design}
We simulate $S = 3$ studies with sample sizes and genotype frequencies of the three ancestral groups of the empirical application, and fix the generative location of the effect, the intercept, and the residual scale at their posterior means from the empirical analysis. We consider three scenarios, one for each heterogeneity scale: $\tau_\beta$ for scenario A, $\tau_\alpha$ for scenario B, and $\tau_\sigma$ for scenario C. In each, the parameter under study is allowed to vary across the studies while the other two are held fixed, so that a single source of heterogeneity is isolated at a time. Within each scenario, the three study-specific values of the parameter under study are placed at the $10$th, $50$th and $90$th percentiles of their generative distribution. For the signed effect $\beta$ they are quantiles of the Student-$t$, so that a sufficiently large heterogeneity places the studies on opposite sides of zero; for the positive parameters $\alpha$ and $\sigma$ they are the quantiles of the $\text{Student-}t_{[0, \infty)}$. Fixing the studies at these quantiles places them symmetrically about the common center and keeps their relative positions identical across scenarios, so that the only quantity changing the spread between studies is the heterogeneity scale $\tau$: a small $\tau$ brings the studies close together, a large $\tau$ pushes them apart, with nothing else varying.

The heterogeneity scale $\tau$ is set at three levels, 10th, 50th and 90th percentiles of its elicited prior: a low, a central, and a high level, with the resulting values reported in Table~\ref{tab:tau_levels}. The extremes are drawn from the prior rather than taken to their formal limits of $\tau \to 0$ and $\tau \to \infty$, so that even the most extreme scenario remains within the range the data deem plausible.

\begin{table}[!ht]
\centering
\caption{Heterogeneity values used as ground truth in the simulated data-generating mechanism. The three levels for each parameter correspond to 10th, 50th and 90th percentiles of its elicited hyperprior.}
\label{tab:tau_levels}
\small
\begin{tabular}{lccc}
\toprule
Scenario & Low & Medium & High \\
\midrule
A ($\tau_\beta$)   & $0.038$ & $0.28$ & $1.23$ \\
B ($\tau_\alpha$)  & $0.049$ & $0.36$ & $1.59$ \\
C ($\tau_\sigma$)  & $0.036$ & $0.26$ & $1.17$ \\
\bottomrule
\end{tabular}
\end{table}

The generative values of $\tau$, reported in Table~\ref{tab:tau_levels}, are used only to simulate the data and are not supplied to the model, which estimates them from the data, as in the empirical analysis. Then, each simulated dataset is analyzed with the same generative model used in the empirical application, in which the heterogeneity parameters $\tau_\beta$, $\tau_\alpha$ and $\tau_\sigma$ are random with their elicited priors. 

We replicate every scenario across $R = 30$ independently simulated datasets. As in the empirical case, posterior sampling is carried out in Stan, running four Hamiltonian Monte Carlo chains of 3000 iterations each: the first 1500 discarded as warmup, for 6000 posterior draws in total. We retain only simulated datasets that meet the convergence criteria. The number retained, $R_{\mathrm{used}}$, is reported in Table~\ref{tab:nused}.

\begin{table}[!ht]
\centering
\caption{Number of replicate datasets retained per scenario and heterogeneity level after the convergence criteria, out of $R=30$ simulated.}
\label{tab:nused}
\begin{tabular}{l ccc}
\toprule
Scenario & Low & Medium & High \\
\midrule
A ($\tau_\beta$) & 23 & 25 & 27 \\
B ($\tau_\alpha$) & 21 & 24 & 24 \\
C ($\tau_\sigma$) & 24 & 22 & 21 \\
\bottomrule
\end{tabular}
\end{table}

In order to assess whether the estimated replicability probabilities reflect the real features of the phenomenon rather than Monte Carlo variation, for each metric and for each level of heterogeneity (low, medium, high), we compute the Monte Carlo standard error. Let $\hat p^{(r)}$ denote the posterior estimate obtained from the $r$-th replicate; we summarize the metric by the across-replicate mean $\bar p = R_{\mathrm{used}}^{-1}\sum_{r}\hat p^{(r)}$ and quantify its precision by the Monte Carlo standard error
\[
  \mathrm{MCSE}(\bar p)=
  \frac{\mathrm{sd}\!\left(\hat p^{(1)},\dots,\hat p^{(R_{\mathrm{used}})}\right)}
       {\sqrt{R_{\mathrm{used}}}}.
\]
Here $\mathrm{MCSE}(\bar p)$ is computed across the $R_{\mathrm{used}}$ simulated datasets, whereas $\mathrm{MCSE}(\hat p)$ in the empirical analysis was computed over the MCMC posterior draws of a single fit.

Point estimates $\bar p$ and their Monte Carlo standard errors $\mathrm{MCSE}(\bar p)$ are reported in Table~\ref{tab:mcse-heterogeneity}. In order to match the threshold used in the empirical analysis, we report the simulation results throughout at the consensus level $k=2$. We mark in bold the entries whose ratio $\bar p/\mathrm{MCSE}(\bar p)$ is below $2$, a situation that can arise either from a very high $\mathrm{MCSE}(\bar p)$ or from a very low $\hat p$.

\begin{table}[!ht]
\centering
\tiny
\caption{Replicability metrics under the simulation scenarios at the consensus level $k = 2$, reported as across-replicate mean $\pm$ $\mathrm{MCSE}(\bar p)$ over the $R_{\mathrm{used}}$ replicate datasets passing the convergence criteria, out of $R=30$ simulated. Values in bold are within two $\mathrm{MCSE}(\bar p)$ of zero ($\bar p/\mathrm{MCSE}(\bar p)<2$); $0$ denotes no occurrences across retained replicates.}
\label{tab:mcse-heterogeneity}
\begin{tabular}{l c c c}
\toprule
Metric & Low & Medium & High \\
\midrule
\multicolumn{4}{@{}l}{\textit{A: Effect heterogeneity ($\tau_\beta$)}} \\
\midrule
$P_{\mathrm{overall},2}$ & $0.99 \pm 0.0013$ & $0.99 \pm 0.0020$ & $0.60 \pm 0.050$ \\
$P_{\mathrm{non\text{-}null},2}$ & $0.15 \pm 0.040$ & $0.16 \pm 0.045$ & $0.15 \pm 0.034$ \\
$P_{\mathrm{pos},2}$ & $0.15 \pm 0.040$ & $0.16 \pm 0.045$ & $0.14 \pm 0.035$ \\
$P_{\mathrm{null},2}$ & $0.85 \pm 0.040$ & $0.83 \pm 0.045$ & $0.45 \pm 0.048$ \\
{\boldmath$P_{\mathrm{neg},2}$} & {\boldmath$0.00077 \pm 0.00040$} & {\boldmath$0.0018 \pm 0.00091$} & {$0.0063 \pm 0.0029$} \\
$P_{\mathrm{cond}\mid O = 1, m = 1}$ & $0.51 \pm 0.044$ & $0.52 \pm 0.054$ & $0.088 \pm 0.039$ \\
$P_{\mathrm{cond}\mid O = 2, m = 1}$ & $0.63 \pm 0.043$ & $0.60 \pm 0.040$ & $0.78 \pm 0.039$ \\
$P_{\mathrm{cond}\mid O = 3, m = 1}$ & $0.45 \pm 0.057$ & $0.40 \pm 0.059$ & $0.22 \pm 0.046$ \\
$P_\beta$ & $0.14 \pm 0.033$ & $0.15 \pm 0.037$ & $0.15 \pm 0.010$ \\
$P_{\mathrm{gen,overall},2}$ & $0.91 \pm 0.013$ & $0.91 \pm 0.012$ & $0.46 \pm 0.046$ \\
$P_{\mathrm{gen,non\text{-}null},2}$ & $0.10 \pm 0.032$ & $0.11 \pm 0.036$ & $0.041 \pm 0.011$ \\
$P_{\mathrm{gen,pos},2}$ & $0.10 \pm 0.032$ & $0.11 \pm 0.036$ & $0.040 \pm 0.011$ \\
$P_{\mathrm{gen,null},2}$ & $0.81 \pm 0.042$ & $0.80 \pm 0.046$ & $0.42 \pm 0.046$ \\
{\boldmath$P_{\mathrm{gen,neg},2}$} & {\boldmath$0.00013 \pm 0.000065$} & {\boldmath$0.00050 \pm 0.00026$} & {\boldmath$0.0016 \pm 0.00080$} \\
$P_{\mathrm{cpos},2}$ & $0.54 \pm 0.044$ & $0.50 \pm 0.053$ & $0.26 \pm 0.048$ \\
$P_{\mathrm{cnull},2}$ & $0.92 \pm 0.022$ & $0.91 \pm 0.025$ & $0.48 \pm 0.050$ \\
$P_{\mathrm{cneg},2}$ & $0.032 \pm 0.012$ & $0.092 \pm 0.043$ & $0.038 \pm 0.015$ \\
\midrule
\multicolumn{4}{@{}l}{\textit{B: Intercept heterogeneity ($\tau_\alpha$)}} \\
\midrule
$P_{\mathrm{overall},2}$ & $0.99 \pm 0.0044$ & $0.98 \pm 0.0042$ & $0.98 \pm 0.0059$ \\
$P_{\mathrm{non\text{-}null},2}$ & $0.34 \pm 0.056$ & $0.40 \pm 0.050$ & $0.41 \pm 0.053$ \\
$P_{\mathrm{pos},2}$ & $0.34 \pm 0.056$ & $0.40 \pm 0.050$ & $0.41 \pm 0.054$ \\
$P_{\mathrm{null},2}$ & $0.65 \pm 0.055$ & $0.58 \pm 0.049$ & $0.57 \pm 0.052$ \\
{\boldmath$P_{\mathrm{neg},2}$} & {\boldmath$0.0018 \pm 0.00090$} & {\boldmath$0.0017 \pm 0.00097$} & {\boldmath$0.0017 \pm 0.00092$} \\
$P_{\mathrm{cond}\mid O = 1, m = 1}$ & $0.77 \pm 0.046$ & $0.80 \pm 0.042$ & $0.79 \pm 0.036$ \\
$P_{\mathrm{cond}\mid O = 2, m = 1}$ & $0.58 \pm 0.046$ & $0.62 \pm 0.040$ & $0.66 \pm 0.043$ \\
$P_{\mathrm{cond}\mid O = 3, m = 1}$ & $0.72 \pm 0.046$ & $0.77 \pm 0.039$ & $0.73 \pm 0.043$ \\
$P_\beta$ & $0.29 \pm 0.047$ & $0.34 \pm 0.043$ & $0.35 \pm 0.046$ \\
$P_{\mathrm{gen,overall},2}$ & $0.85 \pm 0.0096$ & $0.82 \pm 0.0099$ & $0.82 \pm 0.011$ \\
$P_{\mathrm{gen,non\text{-}null},2}$ & $0.25 \pm 0.049$ & $0.28 \pm 0.045$ & $0.30 \pm 0.048$ \\
$P_{\mathrm{gen,pos},2}$ & $0.24 \pm 0.049$ & $0.28 \pm 0.045$ & $0.30 \pm 0.048$ \\
$P_{\mathrm{gen,null},2}$ & $0.60 \pm 0.053$ & $0.53 \pm 0.047$ & $0.52 \pm 0.050$ \\
{\boldmath$P_{\mathrm{gen,neg},2}$} & {$0.00037 \pm 0.00018$} & {\boldmath$0.00035 \pm 0.00022$} & {\boldmath$0.00036 \pm 0.00022$} \\
$P_{\mathrm{cpos},2}$ & $0.75 \pm 0.032$ & $0.78 \pm 0.027$ & $0.78 \pm 0.029$ \\
$P_{\mathrm{cnull},2}$ & $0.80 \pm 0.041$ & $0.75 \pm 0.037$ & $0.75 \pm 0.038$ \\
$P_{\mathrm{cneg},2}$ & $0.063 \pm 0.021$ & $0.036 \pm 0.016$ & $0.042 \pm 0.017$ \\
\midrule
\multicolumn{4}{@{}l}{\textit{C: Residual-scale heterogeneity ($\tau_\sigma$)}} \\
\midrule
$P_{\mathrm{overall},2}$ & $0.99 \pm 0.0037$ & $0.99 \pm 0.0040$ & $0.99 \pm 0.0044$ \\
$P_{\mathrm{non\text{-}null},2}$ & $0.37 \pm 0.049$ & $0.38 \pm 0.055$ & $0.35 \pm 0.066$ \\
$P_{\mathrm{pos},2}$ & $0.36 \pm 0.049$ & $0.38 \pm 0.055$ & $0.35 \pm 0.066$ \\
$P_{\mathrm{null},2}$ & $0.62 \pm 0.048$ & $0.61 \pm 0.054$ & $0.64 \pm 0.065$ \\
{\boldmath$P_{\mathrm{neg},2}$} & {\boldmath$0.0011 \pm 0.00061$} & {\boldmath$0.00086 \pm 0.00047$} & $0$ \\
$P_{\mathrm{cond}\mid O = 1, m = 1}$ & $0.81 \pm 0.032$ & $0.82 \pm 0.034$ & $0.86 \pm 0.024$ \\
$P_{\mathrm{cond}\mid O = 2, m = 1}$ & $0.62 \pm 0.042$ & $0.61 \pm 0.046$ & $0.56 \pm 0.054$ \\
$P_{\mathrm{cond}\mid O = 3, m = 1}$ & $0.75 \pm 0.039$ & $0.74 \pm 0.041$ & $0.65 \pm 0.057$ \\
$P_\beta$ & $0.31 \pm 0.041$ & $0.33 \pm 0.045$ & $0.30 \pm 0.054$ \\
$P_{\mathrm{gen,overall},2}$ & $0.84 \pm 0.010$ & $0.83 \pm 0.012$ & $0.84 \pm 0.014$ \\
$P_{\mathrm{gen,non\text{-}null},2}$ & $0.27 \pm 0.042$ & $0.28 \pm 0.047$ & $0.25 \pm 0.055$ \\
$P_{\mathrm{gen,pos},2}$ & $0.27 \pm 0.042$ & $0.28 \pm 0.047$ & $0.25 \pm 0.055$ \\
$P_{\mathrm{gen,null},2}$ & $0.58 \pm 0.047$ & $0.56 \pm 0.053$ & $0.59 \pm 0.064$ \\
{\boldmath$P_{\mathrm{gen,neg},2}$} & {\boldmath$0.00019 \pm 0.00011$} & {\boldmath$0.00019 \pm 0.00011$} & $0$ \\
$P_{\mathrm{cpos},2}$ & $0.78 \pm 0.026$ & $0.78 \pm 0.029$ & $0.69 \pm 0.044$ \\
$P_{\mathrm{cnull},2}$ & $0.80 \pm 0.033$ & $0.77 \pm 0.043$ & $0.77 \pm 0.047$ \\
$P_{\mathrm{cneg},2}$ & $0.032 \pm 0.012$ & $0.025 \pm 0.011$ & $0$ \\
\bottomrule
\end{tabular}
\end{table}

Across all scenarios the $\mathrm{MCSE}(\bar p)$ values are small (median $\approx 0.04$, maximum $\approx 0.066$). The ratios $\bar p/\mathrm{MCSE}(\bar p)$ below $2$ arise not from large errors but from very low probabilities, that is, from estimates that cannot be distinguished from zero. These are $P_{\mathrm{neg},2}$ and $P_{\mathrm{gen,neg},2}$, the same probabilities, which rest on very few draws. Indeed, in the simulation the study-specific effects are spread around the empirical posterior mean of the generative effect ($0.61$) across the heterogeneity levels, and since this value lies well above $-\varepsilon$, almost none of them fall below $-\varepsilon$; the only exception is the first study at the high level of heterogeneity, as we can see in Table~\ref{tab:beta_scenarios}.

\begin{table}[!ht]
\centering
\caption{Ground-truth heterogeneity levels for the effect (Scenario~A) and the resulting study-specific effects. The three study-specific effects are placed at the 10th, 50th, and 90th percentiles of the generative Student-$t$ distribution, centered at the empirical posterior mean $0.61$, with spread governed by $\tau_\beta$.}
\label{tab:beta_scenarios}
\small
\begin{tabular}{lcc}
\toprule
 & $\tau_\beta$ & $(\beta_1,\beta_2,\beta_3)$ \\
\midrule
Low    & $0.038$ & $(0.55,\ 0.61,\ 0.67)$ \\
Medium & $0.28$  & $(0.16,\ 0.61,\ 1.06)$ \\
High   & $1.23$  & $(-1.40,\ 0.61,\ 2.62)$ \\
\bottomrule
\end{tabular}
\end{table}

\subsection{Results}

\begin{figure}[!ht]
\centering
\includegraphics[width=0.55\textwidth]{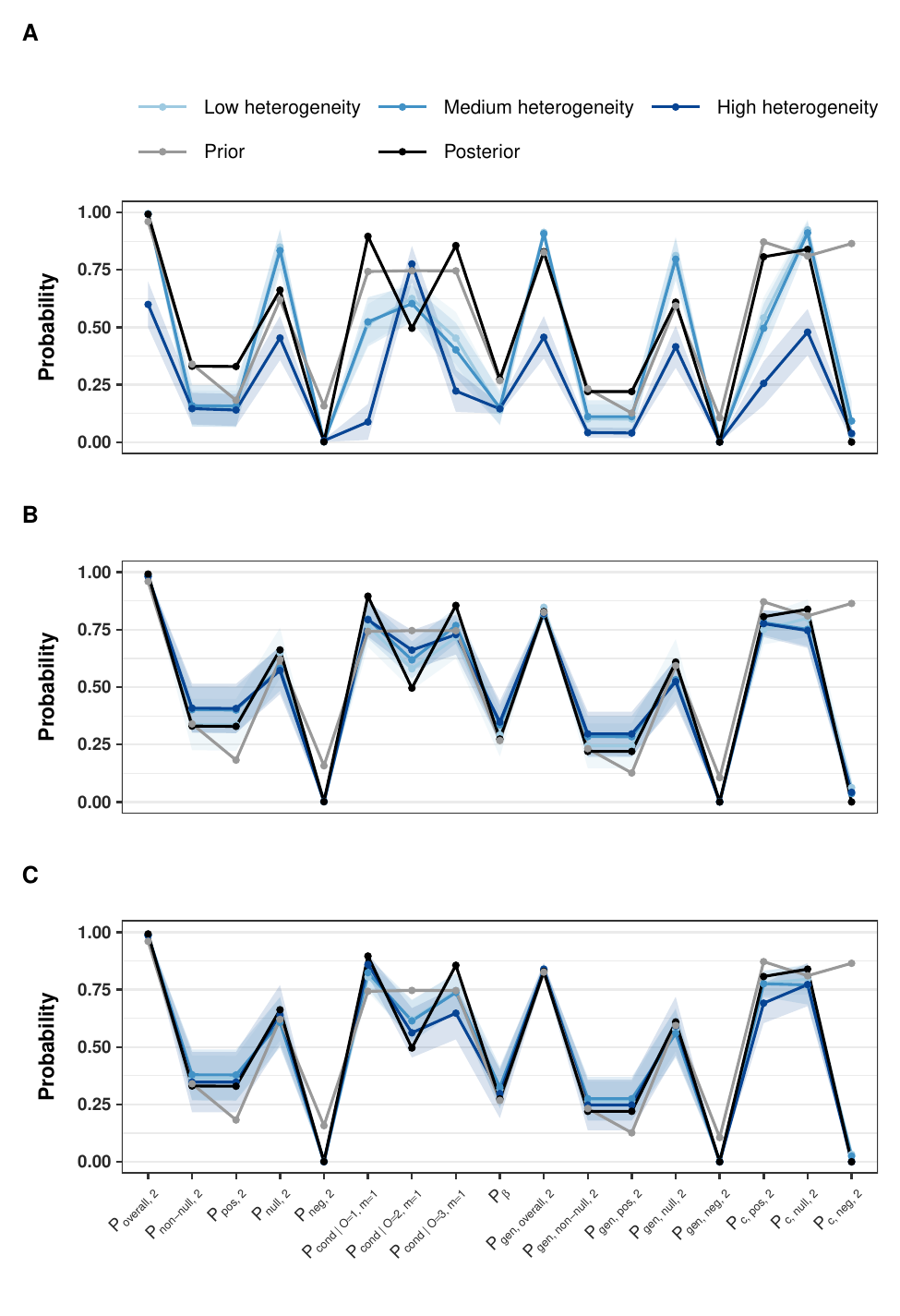}
\caption{Replicability metrics across increasing heterogeneity, at the consensus level $k=2$. Each panel varies one heterogeneity scale, $\tau_\beta$ (A), $\tau_\alpha$ (B), $\tau_\sigma$ (C), holding the others at their empirical values. Colored curves are the simulated metric values; the gray and black lines mark the prior and posterior profiles of the empirical analysis. The bands correspond to the value of the metrics $\bar p \pm\,\mathrm{MCSE}(\bar p)$, reported in Table~\ref{tab:mcse-heterogeneity}.}
\label{fig:simulation}
\end{figure}

Figure~\ref{fig:simulation} shows the replicability metrics for each scenario and for each level of heterogeneity. The top panel illustrates the effect heterogeneity scenario (A), the second one the intercept heterogeneity scenario (B), and the bottom panel the residual heterogeneity scenario (C). The gray line marks the replicability metrics obtained from the prior distribution, from which the posterior profile, represented by the black line, departs; the
bands correspond to $\bar p \pm\,\mathrm{MCSE}(\bar p)$
(Table~\ref{tab:mcse-heterogeneity}).

\subsubsection{Scenario A: Heterogeneity in the Effect}
The first experiment varies $\tau_\beta$, the heterogeneity of the regulatory effect itself, holding the nuisance components at their estimated values.

At the lowest level of $\tau_\beta$, the three study-specific effects are nearly identical $(0.55,\ 0.61,\ 0.67)$, centered at the empirical posterior mean, so the studies agree as closely as the design allows. Because the common center $0.61$ lies within $[-\varepsilon,\varepsilon]$, with $\varepsilon = 0.86$, at low $\tau_\beta$ all three effects are of practical equivalence to the null. The low heterogeneity level is therefore an upper bound for the metrics that place the effect within $[-\varepsilon, \varepsilon]$. The null agreement and the conditional generative null consistency are at their highest, respectively $P_{\text{null},2} = 0.85$ and $P_{\text{cnull},2} = 0.92$, and the generative null consistency is likewise high, $P_{\text{gen,null},2} = 0.81$; the metrics that register an effect of practical relevance are correspondingly low, since no study-specific effect clears the threshold; consequently, the
non-null agreement falls to $P_{\text{non-null},2} = 0.15$. As shown in the top panel of Figure~\ref{fig:simulation}, the empirical values, represented by the black line, sit below the lightest line, which marks the lowest level of heterogeneity, on the null metrics (where the posterior probabilities of the empirical application are $P_{\text{null},2} = 0.66$, $P_{\text{cnull},2} = 0.84$ and $P_{\text{gen,null},2} = 0.61$), yet above it on the directional ones: in the empirical analysis, unlike in the simulated low-heterogeneity scenario, the largest effect is above $\varepsilon = 0.86$ (posterior mean $0.93$) and so the directional probabilities increase.

At the highest level of $\tau_\beta$, $P_{\text{null},2} = 0.45$, $P_{\text{cnull},2} = 0.48$ and $P_{\text{gen,null},2} = 0.42$. The conditional probabilities do not move together but diverge: conditioning on a discovery in the second study, whose effect lies at the center of the distribution, $P_{\text{cond}\mid O = 2, m = 1}$ rises to $0.78$, while conditioning on the first study, whose effect is placed at the lowest percentile, $P_{\text{cond}\mid O = 1, m = 1}$ falls to $0.088$. At high heterogeneity, the three study-specific effects are $(-1.40,\ 0.61,\ 2.62)$: whenever the second study registers a relevant effect, of either sign, a study of the same direction corroborates it: a draw above $\varepsilon$ is matched by the third study-specific effect, whose effect is strongly above $\varepsilon$, and a draw below $-\varepsilon$ by the first study-specific effect. By contrast, when the first study registers its effect strongly below $-\varepsilon$, the only candidates are the central study, of practical equivalence to the null, and the third, strongly above $\varepsilon$, so a second study-specific effect below $-\varepsilon$ almost never appears, and corroboration in the same direction collapses.

\subsubsection{Scenario B: Heterogeneity in the Intercept}
The second and third scenarios vary $\tau_\alpha$ and $\tau_\sigma$ respectively, the heterogeneity of the nuisance parameters, in order to assess the robustness of the conclusions about the replicability of the effect of interest with respect to an increase in the between-study heterogeneity of the nuisance components.

When the level of heterogeneity between the study-specific intercepts changes, the replicability metrics for $\beta$ remain largely unchanged, as shown in the second panel of Figure~\ref{fig:simulation}.

\subsubsection{Scenario C: Heterogeneity in the Residual Scale}
Scenario C varies $\tau_\sigma$, the heterogeneity of the residual scales. In this case as well, the metrics that assess the replicability of the effect of interest remain largely unchanged (bottom panel of Figure~\ref{fig:simulation}). 

Since, apart from the metrics $P_{\mathrm{neg,2}}$ and $P_{\mathrm{gen,neg,2}}$, every metric has $\bar p/\mathrm{MCSE}(\bar p)\ge 2$ (Table~\ref{tab:mcse-heterogeneity}), we can conclude that the values are reliably estimated rather than driven by simulation noise.

Although we have kept the simulation close to the empirical case study, in order to compare the probabilities, the construction is general: any parameter of the model may be perturbed, alone or in combination, and alternative generative mechanisms substituted, with no change to the machinery. 

\section{Discussion}

\subsection{Borrowing Strength versus Pure Replicability}
Since our replicability metrics are computed from draws of the hierarchical model, they also partly measure shrinkage. 
We have discussed why we adopt a hierarchical model: since we are studying the replicability between studies that derive from the same super-population, the assumption that the studies are dependent is reasonable, and shrinkage embodies it, borrowing strength that also stabilizes the metrics for small studies. It further allows us to compute all the generative metrics, and thus to address the important question, discussed above, of the relationship between replicability and meta-analysis. At the same time, the probability metrics obtained through the hierarchical model should be interpreted with caution because they partly measure shrinkage: when one study is much larger than the others, the posterior distributions of the smaller studies might be largely determined by the dominant study and by the prior on $\tau_\beta$. To gauge how much, and to measure replicability more directly, we propose to compute the independence-limit metrics, that isolate each study's own data-level contribution.

\subsection{Identifying the Between-study Heterogeneity with Few Studies}
Specifying a prior for $\tau_\beta$, the parameter that governs how strongly the study-specific effects $\beta_s$ are pooled toward the generative effect $\beta$, is not straightforward, especially when only a few studies are available: with so little evidence, it is hard to judge what a plausible degree of between-study heterogeneity would be. Three broad approaches are available, following the classification of \citet{spiegelhalter2004bayesian}. The first is elicitation: because the spread of effects implied by $\tau_\beta$ has a concrete reading, a researcher can calibrate a prior by asking how large a discrepancy between two comparable studies could be at most \citep{spiegelhalter2004bayesian, o2019expert}. The second summarizes past evidence: when previous analyses in the same domain exist, the distribution of their estimated heterogeneities can itself be analyzed to yield an empirically grounded prior \citep{turner2012predicting}. The third relies on default priors, including inverse-gamma, half-$t$, and half-Cauchy families; see \citet{gelman2006prior, polson2012half, chung2013nondegenerate} for discussion of their behavior.

Our empirical application sits within the second strategy, but in an unusual form. For a given SNP--transcription factor pair, there is typically no prior meta-analysis from which to read off a plausible heterogeneity, so this route is unavailable in its literal form. What the genome offers instead is a large universe of biologically analogous regulatory relationships: many other pairs whose effects have been measured across studies, and whose between-study variability can be characterized even when a specific pair has not been studied enough to support a meta-analysis. The empirical Bayes prior for $\tau_\beta$ is built from this universe.

First, with few studies, the variance-parameter estimates of hierarchical linear models frequently collapse to zero at the boundary of the parameter space; placing a proper prior on $\tau$, the standard Bayesian counterpart to penalized-likelihood corrections \citep{chung2013nondegenerate}, prevents this collapse even when only a handful of studies is available.

Second, building the prior from the pool of pairs is what keeps the prior itself from degenerating to a spike at zero, even though $\tau_\beta$, $\tau_\alpha$ and $\tau_\sigma$ are each computed on a single pair. Seen this way, pooling across many biological pairs supplies the information that a single pair, with few studies, cannot provide on its own. This construction is an empirical-Bayes device \citep{Statistics:112828}, and so it carries the trade-offs characteristic of that approach. Although this comes at the cost that the elicited prior is estimated from the same
donors used in the analysis, we still prefer a pool of similar biological pairs to standard generic priors, because genomics provides a large set of comparable relationships from which plausible, realistic values for the effect of interest can be obtained. It is worth noting that the pool is formed from pairs other than the target, so the hyperparameters are informed by comparable relationships rather than by the target pair's own regression.

\subsection{Assessing Replicability without a Prespecified Threshold}
The proposed replicability metrics rely on the prespecified threshold $\varepsilon$ that defines the region of practical relevance. Trying different thresholds $\varepsilon$ requires no refitting: the metrics are recomputed from the existing posterior draws.

Ideally, the threshold should be set independently of the data, on the basis of scientific judgment. Yet, in truly novel contexts where such information is absent, the framework can be read through summaries without a prespecified boundary. One could examine the posterior distribution of the difference between the study-specific effects, for example, between the largest and smallest ones, or summarize how far the posterior distributions of the study-specific effects overlap. 

We kept the main analysis anchored to the single threshold $\varepsilon$ for clarity, but the code to compute these threshold-free summaries is available in the companion repository (\url{https://github.com/esteralongi/The-Quantification-Of-Replicability}).

\subsection{Replicability across Many Studies}
Our development has kept the number of studies $S$ small. However, it is worth asking how the framework behaves as $S$ grows. The replicability questions that we discussed scale naturally. With the consensus level $k$, we can already measure replicability within subsets of studies rather than across all of them, which helps especially when the collection grows. Inference on the generative effect becomes more reliable as more studies inform it; and the predictive questions persist in both directions, since the accumulated evidence can be used prospectively to forecast a future study, or in a leave-one-out setting, holding out each study in turn to assess retrospective replicability across the collection.

However, the joint metrics presuppose that the studies are exchangeable members of one super-population, an assumption more easily violated when $S$ is large, since the collection may then contain clusters of studies. Thus, it could be worth assessing exchangeability jointly with the replicability analysis; this idea has been formalized in related  settings through Bayesian nonparametric models that embed clustering and model comparison in a single joint framework \citep{10.1214/14-AOAS798}. In our framework, one might fit the generative model on all studies except one and check how well the held-out study agrees with it.

Even within a subset of studies, one could assess replicability pairwise, two studies at a time, as in the Reproducibility Project: Cancer Biology \citep{Errington2021}, rather than jointly. The two are not interchangeable, and the choice can change the conclusions. \citet{ganzfried2013curatedovariandata} illustrate this difference concretely with the prognostic role of \textit{CXCL12} expression in ovarian cancer: the association with survival emerged from the combined analysis of many studies, in only a few of which the gene was individually a significant predictor. A pairwise assessment would register very few successes, since most pairs contain no two individually significant studies. However, the combined analysis finds a coherent signal precisely because it accumulates evidence that individual studies, each too small on its own, cannot provide alone, even if it can be driven by a minority of studies. Therefore, pairwise and joint assessments answer different questions, and the choice between them should be based on the scientific goal.

Together, these considerations suggest a richer extension of our framework to the many-study setting. However, a full development lies beyond the scope of this article, whose primary aim is to introduce a new conceptualization of replicability; we leave it to future work.

\subsection{Extension to More Parameters}
We have illustrated our work through a linear regression in which a single coefficient $\beta$ is the object of interest, in order to present the framework as clearly as possible. However, neither the model nor the parameter is essential. The framework rests only on a generative model and on joint posterior probabilities, and nothing requires a single coefficient: a model with additional covariates and several coefficients of interest is still handled, because replicability is still assessed through the same posterior probabilities over a region of practical relevance, applied to each coefficient. When several coefficients are of joint interest, that region can be applied to each of them marginally or specified jointly in the parameter space.

\subsection{The File Drawer Problem}
``The journals are filled with the 5\% of the studies that show Type I errors, while the file drawers back at the lab are filled with the 95\% of the studies that show nonsignificant results.'' \citep{rosenthal1979file}. As early as in 1979, Robert Rosenthal discussed a well-known problem that remains open today: a collection of published studies may represent a selected subset of all the studies actually carried out. This phenomenon, now commonly called \emph{publication bias}, can distort the distribution of the observed study-specific effects and, consequently, the assessment of replicability based on them.

In our empirical application, there is no publication bias in the choice of the studies: the studies are observed independently of the parameters that generated them. Since we defined the studies by taking the ancestral groups from the entire GTEx dataset, they are not selected based on some criterion of significance of their effects.

When feasible, preregistered multi-lab studies, including multi-lab Registered Reports \citep{nosek2014registered}, provide a natural prospective setting for our framework. In Registered Reports, the research questions, study design, and analysis plan are peer-reviewed before data collection, and the publication decision is made primarily on the basis of the proposed questions and methods rather than on the direction or statistical significance of the results. This substantially reduces the outcome-dependent selection of studies underlying publication bias. Therefore, although such coordinated designs are not always feasible and do not replace retrospective analyses of existing evidence, they offer a promising direction for assessing replicability transparently from the outset.

If, instead, one wanted to assess the replicability of effects from already published studies, the probability that a study is observed may depend on its results, and so it would be necessary to take publication bias into account. In the meta-analysis literature, several methods for correcting this bias have been proposed; \citet{haaf2020conventional} provides a review of these methods.

In the framework of replicability that we proposed here, one could leverage the fact that a generative model fully specifies the process that produces the studies, and can therefore be extended by introducing an explicit selection mechanism, as \citet{bayarri1987bayesian} proposed. By modifying the generative model so that it also generates the data that end up in the file drawer because of their nonsignificant results, one could then apply the replicability framework we proposed, computing the proposed metrics with respect to the latent population of conducted studies rather than only the observed subset. This would make it possible to distinguish replicability within the available evidence from the replicability expected across the broader process that generated the studies.

\subsection{Broader Applicability}
Although we have illustrated the framework through an application in genomics, its architecture is independent of that specific domain: it applies to any field that combines evidence across studies sharing a common super-population, for example through a hierarchical structure. Multi-site clinical trials, large-scale replication projects in psychology and the social sciences, and multi-laboratory experiments all share study-specific effects related to one another through a common generative effect. In each of these settings, the metrics would quantify the replicability of the effects of interest, revealing not only whether the studies replicate but also why, by examining the joint and conditional probabilities linking the generative effect, the study-specific effects, and their heterogeneity. 
The framework provides a systematic assessment of replicability before broad scientific claims are drawn. When replicability is lacking, it further makes it possible to investigate its source and, where useful, to inform the design of a new study.

\section*{Acknowledgments}
We would like to express our deepest gratitude to Branden Fitelson for his invaluable advice and deeply inspiring approach to scientific inquiry, and to Filippo Gambarota for engaging and insightful discussions about replicability.

\bibliographystyle{plainnat}
\bibliography{bibliography}

\section*{Supplementary Material}\label{supp:mcse}
Tables~\ref{tab:mcse-indep}--\ref{tab:mcse-held} report the Monte Carlo diagnostics: the value of each metric under the prior alone $p_{\mathrm{prior}}$, the posterior estimate $\hat p$, the number of draws meeting the event $T^{*}$ and the number of draws $T$ (for conditional metrics, the number satisfying the conditioning event), the effective sample size ESS, and $\mathrm{MCSE}(\hat p)=\sqrt{\hat p(1-\hat p)/\mathrm{ESS}}$. Events not observed in any of the $6000$ draws are reported as the rule-of-three upper bound $3/T$ ($<$); conditional estimates with fewer than $30$ conditioning draws are marked with $\dagger$.

\begin{table}[!ht]\centering\footnotesize\setlength{\tabcolsep}{4pt}
\caption{Independence limit ($k=2$/$k=3$).}\label{tab:mcse-indep}
\begin{tabular}{l r r r r r r}
\toprule
Metric & $p_{\mathrm{prior}}$ & $\hat p$ & $T^*$ & $T$ & ESS & $\mathrm{MCSE}(\hat p)$ \\ \midrule
$P_{\mathrm{overall},3}$ & 0.21 & 0.066 & 395 & 6000 & 5478 & 0.0033 \\
$P_{\mathrm{overall},2}$ & 0.91 & 0.84 & 5048 & 6000 & 5040 & 0.0052 \\
$P_{\mathrm{non\text{-}null},3}$ & 0.21 & 0.012 & 70 & 6000 & 5050 & 0.0015 \\
$P_{\mathrm{non\text{-}null},2}$ & 0.90 & 0.41 & 2443 & 6000 & 4530 & 0.0073 \\
$P_{\mathrm{pos},3}$ & 0.11 & 0.012 & 70 & 6000 & 5050 & 0.0015 \\
$P_{\mathrm{neg},3}$ & 0.10 & $<0.0005$ & 0 & 6000 & --- & --- \\
$P_{\mathrm{null},3}$ & 0.00034 & 0.054 & 325 & 6000 & 5250 & 0.0031 \\
$P_{\mathrm{pos},2}$ & 0.45 & 0.39 & 2343 & 6000 & 4564 & 0.0072 \\
$P_{\mathrm{neg},2}$ & 0.45 & 0.017 & 100 & 6000 & 5262 & 0.0018 \\
$P_{\mathrm{null},2}$ & 0.013 & 0.43 & 2605 & 6000 & 4748 & 0.0072 \\
$P_{\mathrm{cond}\mid O=1,m=1}$ & 0.72 & 0.22 & 285 & 1314 & 1314 & 0.011 \\
$P_{\mathrm{cond}\mid O=2,m=1}$ & 0.72 & 0.45 & 2333 & 5159 & 5159 & 0.0069 \\
$P_{\mathrm{cond}\mid O=3,m=1}$ & 0.72 & 0.73 & 2338 & 3200 & 3200 & 0.0078 \\
\bottomrule
\end{tabular}
\end{table}

\begin{table}[!ht]\centering\footnotesize\setlength{\tabcolsep}{4pt}
\caption{Hierarchical model; columns as in Table~\ref{tab:mcse-indep}.}\label{tab:mcse-hier}
\begin{tabular}{l r r r r r r}
\toprule
Metric & $p_{\mathrm{prior}}$ & $\hat p$ & $T^*$ & $T$ & ESS & $\mathrm{MCSE}(\hat p)$ \\ \midrule
$P_{\mathrm{overall},3}$ & 0.62 & 0.43 & 2604 & 6000 & 2948 & 0.0091 \\
$P_{\mathrm{overall},2}$ & 0.96 & 0.99 & 5954 & 6000 & 5690 & 0.0012 \\
$P_{\mathrm{non\text{-}null},3}$ & 0.19 & 0.15 & 873 & 6000 & 2978 & 0.0065 \\
$P_{\mathrm{non\text{-}null},2}$ & 0.34 & 0.33 & 1982 & 6000 & 3386 & 0.0081 \\
$P_{\mathrm{pos},3}$ & 0.10 & 0.15 & 873 & 6000 & 2978 & 0.0065 \\
$P_{\mathrm{neg},3}$ & 0.087 & $<0.0005$ & 0 & 6000 & --- & --- \\
$P_{\mathrm{null},3}$ & 0.44 & 0.29 & 1731 & 6000 & 2557 & 0.0090 \\
$P_{\mathrm{pos},2}$ & 0.18 & 0.33 & 1975 & 6000 & 3345 & 0.0081 \\
$P_{\mathrm{neg},2}$ & 0.16 & 0.0012 & 7 & 6000 & 6000 & 0.00044 \\
$P_{\mathrm{null},2}$ & 0.62 & 0.66 & 3972 & 6000 & 3404 & 0.0081 \\
$P_{\mathrm{cond}\mid O=1,m=1}$ & 0.74 & 0.90 & 1266 & 1413 & 1413 & 0.0081 \\
$P_{\mathrm{cond}\mid O=2,m=1}$ & 0.74 & 0.50 & 1885 & 3802 & 3802 & 0.0081 \\
$P_{\mathrm{cond}\mid O=3,m=1}$ & 0.74 & 0.86 & 1686 & 1970 & 1970 & 0.0079 \\
$P_\beta$ & 0.27 & 0.27 & 1645 & 6000 & 2818 & 0.0084 \\
$P_{\mathrm{gen,overall},3}$ & 0.60 & 0.42 & 2498 & 6000 & 2971 & 0.0090 \\
$P_{\mathrm{gen,overall},2}$ & 0.83 & 0.83 & 4974 & 6000 & 4063 & 0.0059 \\
$P_{\mathrm{gen,non\text{-}null},3}$ & 0.17 & 0.14 & 812 & 6000 & 3003 & 0.0062 \\
$P_{\mathrm{gen,non\text{-}null},2}$ & 0.23 & 0.22 & 1320 & 6000 & 2931 & 0.0077 \\
$P_{\mathrm{gen,pos},3}$ & 0.092 & 0.14 & 812 & 6000 & 3003 & 0.0062 \\
$P_{\mathrm{gen,neg},3}$ & 0.079 & $<0.0005$ & 0 & 6000 & --- & --- \\
$P_{\mathrm{gen,null},3}$ & 0.43 & 0.28 & 1686 & 6000 & 2561 & 0.0089 \\
$P_{\mathrm{gen,pos},2}$ & 0.13 & 0.22 & 1320 & 6000 & 2931 & 0.0077 \\
$P_{\mathrm{gen,neg},2}$ & 0.11 & $<0.0005$ & 0 & 6000 & --- & --- \\
$P_{\mathrm{gen,null},2}$ & 0.59 & 0.61 & 3654 & 6000 & 3374 & 0.0084 \\
$P_{\mathrm{cpos},2}$ & 0.87 & 0.81 & 1320 & 1636 & 1636 & 0.0098 \\
$P_{\mathrm{cneg},2}$ & 0.86 & $<0.33^{\dagger}$ & 0 & 9 & --- & --- \\
$P_{\mathrm{cnull},2}$ & 0.81 & 0.84 & 3654 & 4355 & 4355 & 0.0056 \\
\bottomrule
\end{tabular}
\end{table}

\begin{table}[!ht]\centering\footnotesize\setlength{\tabcolsep}{4pt}
\caption{Predictive (retrospective) against the left-out African ancestral group.}\label{tab:mcse-held}
\begin{tabular}{l r r r r r r}
\toprule
Metric & $p_{\mathrm{prior}}$ & $\hat p$ & $T^*$ & $T$ & ESS & $\mathrm{MCSE}(\hat p)$ \\ \midrule
$P_{\mathrm{overall},\ell}$ & 0.22 & 0.36 & 2165 & 6000 & 6000 & 0.0062 \\
$P_{\mathrm{null},\ell}$ & 0.040 & 0.33 & 2006 & 6000 & 6000 & 0.0061 \\
$P_{\mathrm{non\text{-}null},\ell}$ & 0.18 & 0.027 & 159 & 6000 & 6000 & 0.0021 \\
$P_{\mathrm{pos},\ell}$ & 0.094 & 0.019 & 113 & 6000 & 6000 & 0.0018 \\
$P_{\mathrm{neg},\ell}$ & 0.084 & 0.0077 & 46 & 6000 & 6000 & 0.0011 \\
\bottomrule
\end{tabular}
\end{table}

\end{document}